\documentclass{aastex63}

\usepackage{rotating}
\usepackage{multirow,makecell,stackengine}

\graphicspath{{./}{Figures/}}
\begin{document}

\title{Neptune's Spatial Brightness Temperature Variations from the VLA and ALMA}

\correspondingauthor{Joshua Tollefson}
\email{jtollefs@berkeley.edu}

\author{Joshua Tollefson}
\affil{Astronomy Department, University of California, Berkeley; Berkeley CA, 94720, USA}

\author{Imke de Pater}
\affiliation{Astronomy Department, University of California, Berkeley; Berkeley CA, 94720, USA}
\affiliation{Faculty of Aerospace Engineering, Delft University of Technology, Delft 2629 HS, The Netherlands}

\author{Edward M. Molter}
\affiliation{Astronomy Department, University of California, Berkeley; Berkeley CA, 94720, USA}

\author{Robert J. Sault}
\affil{School of Physics, University of Melbourne, Victoria, Australia}

\author{Bryan J. Butler}
\affil{National Radio Astronomy Observatory, Socorro, NM, United States}

\author{Statia Luszcz-Cook}
\affiliation{Department of Astronomy, Columbia University, Pupin Hall, 538 West 120th Street, New York City, NY 10027
}
\affiliation{Astrophysics Department, American Museum of Natural History, Central Park West at 79th Street, New York, NY 10024, USA}

\author{David DeBoer}
\affiliation{Astronomy Department, University of California, Berkeley; Berkeley CA, 94720, USA}



\begin{abstract}
We present spatially resolved ($0.1'' - 1.0''$) radio maps of Neptune taken from the Very Large Array and Atacama Large Submillimeter/Millimeter Array between $2015-2017$. Combined, these observations probe from just below the main methane cloud deck at $\sim 1$ bar down to the NH$_4$SH cloud at $\sim50$ bar. Prominent latitudinal variations in the brightness temperature are seen across the disk. Depending on wavelength, the south polar region is $5-40$ K brighter than the mid-latitudes and northern equatorial region. We use radiative transfer modeling coupled to Markov Chain Monte Carlo methods to retrieve H$_2$S, NH$_3$, and CH$_4$ abundance profiles across the disk, though only strong constraints can be made for H$_2$S. Below all cloud formation, the data are well fit by $53.8^{+18.9}_{-13.4}\times$ and $3.9^{+2.1}_{-3.1}\times$ protosolar enrichment in the H$_2$S and NH$_3$ abundances, respectively, assuming a dry adiabat. Models in which the radio-cold mid-latitudes and northern equatorial region are supersaturated in H$_2$S are statistically favored over models following strict thermochemical equilibrium. H$_2$S is more abundant at the equatorial region than at the poles, indicative of strong, persistent global circulation. Our results imply that Neptune's sulfur-to-nitrogen ratio exceeds unity as H$_2$S is more abundant than NH$_3$ in every retrieval. The absence of NH$_3$ above 50 bar can be explained either by partial dissolution of NH$_3$ in an ionic ocean at GPa pressures or by a planet formation scenario in which hydrated clathrates preferentially delivered sulfur rather than nitrogen onto planetesimals, or a combination of these hypotheses.
\end{abstract}

\section{Introduction}
\label{S:1}

Neptune is the prototypical `ice giant': a giant planet composed mainly of elements heavier than hydrogen and helium by mass, such as oxygen, nitrogen, carbon, and sulfur. Within the cold environment of Neptune, the products expected to form from these elements in the observable atmosphere include: H$_2$O, NH$_3$, CH$_4$, and H$_2$S. These molecules provide clues to the bulk composition of the planet and so constraining their abundances is crucial for understanding Neptune's formation and thermal history. 

Bright CH$_4$ clouds and aerosol hazes pervade throughout Neptune's upper atmosphere. Optical and near-infrared wavelengths are sensitive to these components, limiting views in the atmosphere to pressures less than $\sim1$ bar. H$_2$S and NH$_3$ condense at higher temperatures and pressures than CH$_4$, meaning optical and near-infrared observations are blind to their deep abundances. Radio wavelengths probe beyond these shallow features, resolving the structure of Neptune's atmosphere down to $\sim50$ bar. Thus, NH$_3$ and H$_2$S profiles can be constructed by inverting radio spectra. While early radio observations could only obtain disk-averaged measurements of Neptune, the observed high brightness temperatures longward of 10 cm required NH$_3$, a prominent microwave absorber, to be significantly depleted \citep{dePater1985}. This is possible if an NH$_4$SH cloud forms at $\sim50$ bar and if the H$_2$S abundance exceeds that of NH$_3$, resulting in the complete removal of NH$_3$ during the cloud's formation. In such an atmosphere, an H$_2$S abundance between $30-60\times$ solar and an NH$_3$ abundance of $\sim1\times$ solar are required to fit the disk-averaged data \citep{dePater1991, DeBoer1996}. Indeed, the detection of H$_2$S spectral features near 1.58 $\mu$m in the tropospheres of the ice giants implies that their deep bulk S/N ratio is greater than one \citep{Irwin2018, Irwin2019a}.

\citet{dePater2014} presented centimeter maps of Neptune from 2003, finding that the disk-averaged spectrum agreed with the abundances obtained from earlier radio observations. In addition, they found that the bright south polar cap must be significantly depleted in H$_2$S down to $\sim40$ bar in order to match the observed brightness temperature at wavelengths of $0.7-6.0$ cm. However, this study did not investigate brightness variations at other latitudes, as the sensitivity and resolution were not good enough to detect significant variations apart from those in the south polar cap.

In 2011, an upgrade of the VLA was completed. This expansion improved the continuum sensitivity by 5-to-20-fold and increased the wavelength coverage and bandwidth. This prompted a program to reobserve Neptune at centimeter wavelengths. The resulting maps, presented in this paper, show clear brightness temperature variations across the disk akin to that seen in millimeter maps produced from the Atacama Large Submillimeter/Millimeter Array (ALMA) \citep{Tollefson2019}. The mm and cm probe between $\sim1-50$ bar on Neptune, meaning the most complete picture of Neptune's upper atmosphere to date can be reconstructed by synthesizing these data. As NASA and ESA debate the merits of a next-decade ice giant mission, a firm handle on the uncertainties in the composition and dynamics of Neptune's upper atmosphere are critical \citep{Hofstadter2019}. Just how well are the quantities of N, S, C, P, and O constrained from ground-based observations, in-situ, and orbital measurements, and how does the amount and distribution of condensibles affect the observed atmospheric dynamics \citep{Atreya2019, Hueso2019, Fletcher2020}? Do our uncertainties on these elements necessitate an instrument like \textit{Juno}'s Microwave Radiometer (MWR) \citep{Janssen2017} on a future spacecraft to Neptune \citep{Rymer2020}? What bands and configurations would be most useful for Neptune atmospheric science with the next generation VLA (ngVLA) and could it replace an MWR equivalent?

This paper is organized as follows. First, we present longitudinal-smeared maps of Neptune taken with the expanded VLA in 2015  between  $0.9-9.7$ cm (Section 2). Next, we outline the structure of Neptune's upper atmosphere and the free parameters used in our modeling (Section 3). We then combine these new VLA maps with 2003 VLA and ALMA observations of Neptune and use a Markov chain Monte Carlo (MCMC) implementation of the radiative transfer code Radio BErkeley Atmospheric Radiative transfer (Radio-BEAR) to obtain retrievals for the abundance profiles of Neptune's condensibles (Section 4). Finally, we compare our findings to prior results (Section 5) and end with a summary of key takeaways (Section 6).


\section{Observations}

\subsection{Data}

This work makes use of two primary data sets: 1) ALMA millimeter observations of Neptune taken from $2016-2017$, described in Section 2 of \citet{Tollefson2019}; 2) centimeter observations of Neptune taken with the upgraded VLA taken in 2015, described below.  

We observed Neptune with the expanded VLA, an interferometer located near Socorro, New Mexico, on 1 and 2 September, 2015. Maps were obtained at wavelengths of 0.9 cm (Band Ka), 2.0 cm (Band Ku), 3.0 cm (Band X), 5.1 cm (Band C), and 9.7 cm (Band S). The VLA consists of 27 antennas grouped into three arms of nine antennae to form a `Y'-shape. Every four months, the configuration is changed by moving the antennae along tracks. The `A' configuration is the VLA's most extended; the length of each arm is $\sim 21$ km, forming a maximum baseline of $\sim 36$ km. The maximum baseline is inversely related to the angular resolution, i.e. beam size, meaning variations across Neptune's disk are most distinct in the A configuration. The observation setup strongly impacts the shape of the beam. The symmetry of the VLA three-armed track, Neptune's low declination at the time of the observation, and extended time on source causing the uv-plane sampling to fill out due to Earth’s rotation all contribute to the beams' shapes.\footnote{Further information on the various VLA configurations can be found in the VLA guides for proposers, Section 2: \url{https://science.nrao.edu/facilities/vla/docs/manuals/propvla/referencemanual-all-pages}. An introduction to radio astronomy fundamentals is available online via the Socorro Imaging Synthesis Workshop at \url{https://science.nrao.edu/science/meetings/2018/16th-synthesis-imaging-workshop/16th-synthesis-imaging-workshop-lectures}} Neptune was observed on two days, each for 7 hours divided into many 5 minute scans which rotate through all wavelength bands. Thus, the degree of longitudinal smearing is high in the resulting maps as we observe Neptune throughout nearly an entire 16.11 hr rotation period \citep{Warwick1989}. Table \ref{table:vla-obs} lists a summary of our observations. 

We supplement these observations with VLA maps in 2003 from \citet{dePater2014} in order to model Neptune's disk-average temperature. These observations were taken in 5 bands for a total of 8 hours in each wavelength, including calibrators. Three of these bands (2 cm, Ku; 3.6 cm, X; 6 cm, C) were also taken in the `A' configuration while the other 2 (0.7 cm, Q; 1.3 cm, K) were taken in the `BnA' configuration, which is a hybrid of the B and A configurations, and was ideal for imaging Neptune at low declination.

\begin{table}
\caption{Summary of VLA observations. }
\begin{tabular}{llllllc}
\hline
 Wavelength (cm) & Frequency (GHz) & Band & Beam Size (arcsec$^2$) & Resolution (km$^2$)$^a$ & Time on Source (min)\\
 \hline
\hline
0.9 & 32.958 & Ka & $0.12 \times 0.12$ & $2,486\times2,486$ & 156.0 \\
2.0 & 14.880 & Ku & $0.25 \times 0.25$ & $5,179\times5,179$ & 136.8\\
3.0 & 9.869 & X & $0.35 \times 0.26$ & $7,251\times5,387$ & 139.8 \\
5.1 & 5.861 & C & $0.56 \times 0.45$ & $11,603\times9,324$ & 82.8 \\
9.7 & 3.096 & S & $0.98 \times 0.79$ & $20,305\times16,369$ & 82.2 \\
 \hline
\end{tabular}\label{table:vla-obs}

\footnotesize{$^a$ Resolution at sub-observer location, assuming equatorial and polar radii of 24,766 km and 24,323 km, respectively \citep{Lindal1992}.}

\end{table}

\subsection{Calibration and Imaging}

The VLA observations were loaded from the NRAO data archive and converted for use in the MIRIAD software package \citep{Sault1995}. Flagging, calibration, and imaging were performed within MIRIAD. The calibrators used were 3C48 (flux density) and J2246-1206 (phase). In addition, self-calibration was performed to correct for short-term variability in the phases caused by fast atmospheric fluctuations. The self-calibration model used was a limb-darkened disk that best matched the observations. The limb-darkened profile is represented by the peak brightness temperature $T_b$ multiplied by $\cos^k \theta $, where $\theta$ is the emission angle and $k$ is a limb-darkening constant. Values for $T_b$ and $k$ were found at each wavelength such that the difference between the limb-darkened model disk and observations were minimized. 

\begin{sidewaysfigure}
\centering
  \includegraphics[width=\linewidth]{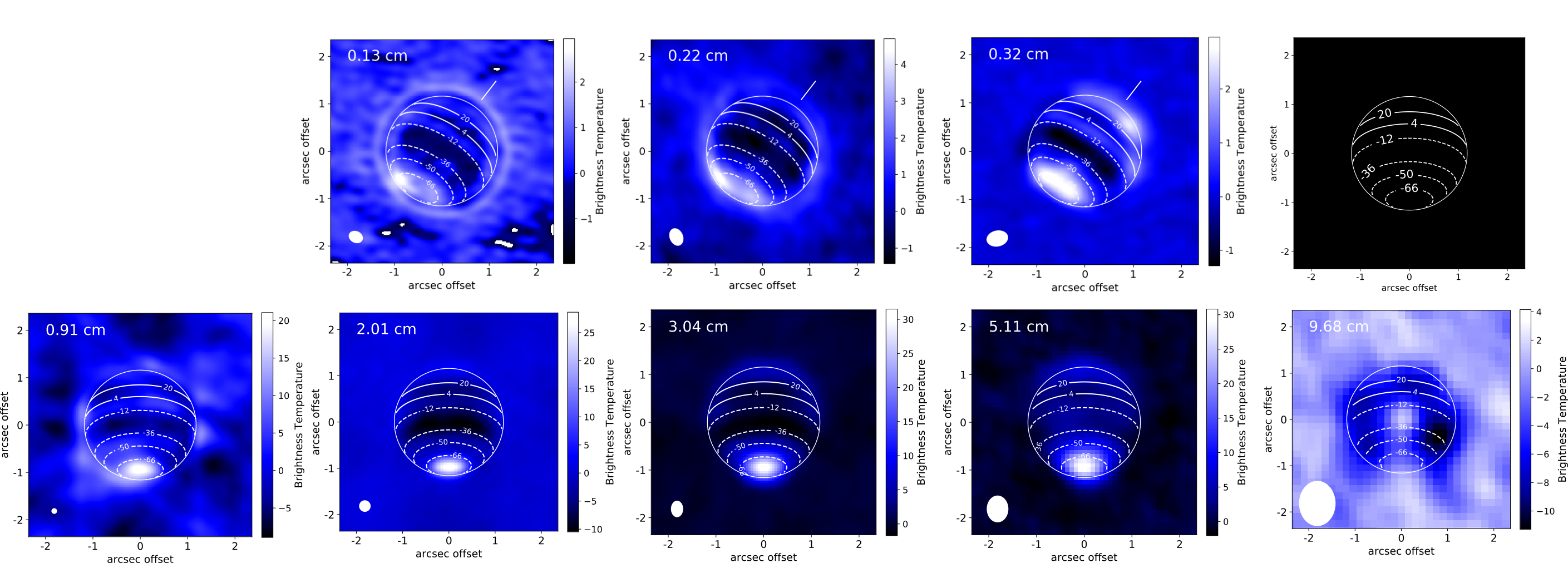}
\caption{Longitude-smeared maps of Neptune taken with ALMA (top row) and the 
upgraded VLA (bottom row). The color scale has been chosen to enhance the brightness contrasts across the disk. All maps are residuals, where a uniform limb-darkened model has been subtracted to highlight the temperature contrasts between latitudes. The north pole is indicated by a white line in the ALMA maps; the VLA maps are rotated so that the north pole is pointing upward. Contour lines delineate the latitude transitions between bands, with a grid over a blank disk shown in the upper right for clarity. Neptune's disk is outlined with a white ellipse. The FWHM of the beam is indicated in white in the bottom left of each map.}
\label{fig:vla-maps}
\end{sidewaysfigure}

\begin{figure}
\centering
  \includegraphics[width=0.5\linewidth]{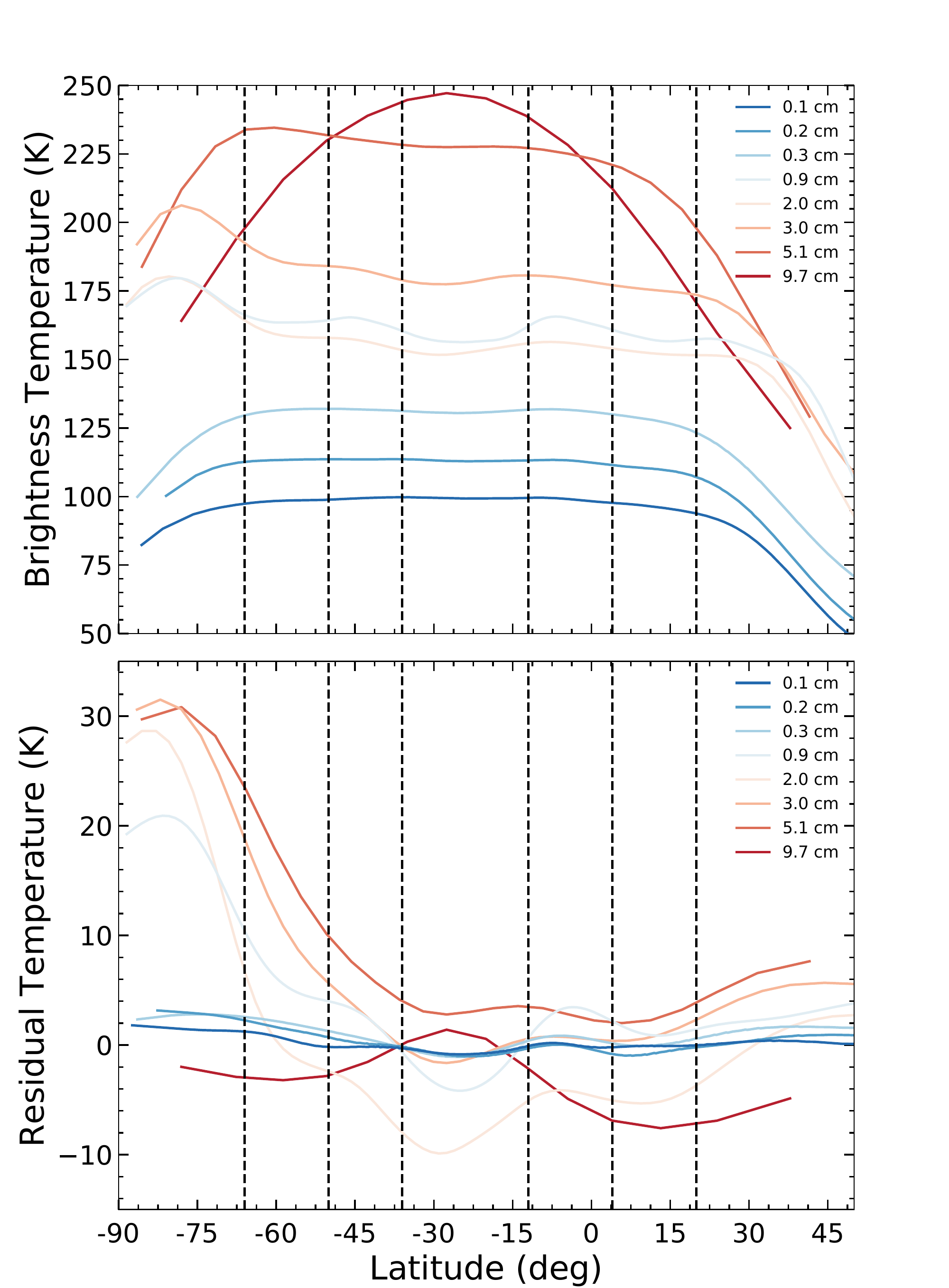}
\caption{Neptune's observed brightness temperature (top panel) and residual temperature (bottom panel) versus latitude along the sub-observer longitude for each wavelength band (colored lines). The residual temperatures are taken from the residual maps shown in Fig. \ref{fig:vla-maps} and are obtained by subtracting the observed brightness temperatures from best-fitting uniform limb-darkened disks, as described in in Section 2.2 Dashed black lines delineate regions where brightness temperature variations are evident and define the latitude bins used in our modeling. Lines may not extend to $90^{\circ}$S due to limits in the resolution of the observation. The bump in the residual temperature at 9.7 cm is due to the poor resolution, as the large beam size impacts the limb-darkened fit.}
\label{fig:temp-slices}
\end{figure}

Our final longitudinally-smeared maps after subtracting the limb-darkened disk of Neptune are shown in Figure \ref{fig:vla-maps}. While the images are `longitudinally-smeared', pixel-to-pixel variations are still present on the disk due to the presence of artifacts on the order of the beam size. Figure \ref{fig:temp-slices} plots the observed brightness temperature and residuals versus latitude along the central meridian of each disk. We divide Neptune into seven regions where the transitions in the brightness temperature variations are apparent: $90^{\circ}$S$-66^{\circ}$S, $66^{\circ}$S$-50^{\circ}$S, $50^{\circ}$S$-36^{\circ}$S, $36^{\circ}$S$-12^{\circ}$S, $12^{\circ}$S$-4^{\circ}$N, $4^{\circ}$N$-20^{\circ}$N, $20^{\circ}$N$-50^{\circ}$N.

\subsection{Error estimation}

 The random noise in our VLA maps is calculated by averaging over four regions of the sky with boxes equal to the diameter of Neptune and taking the root-mean-square (RMS). RMS values range from $0.4-2.0$K and are similar to the RMS at a given latitude within the disk of a residual map. Table \ref{table:vla-disk-avg} lists our estimated errors in each band. The RMS does not include systematic effects, such as errors in the bandpass or flux calibration. Uncertainties in the flux density are estimated at $5\%$ or less in each band so we use this as a conservative estimate for the absolute error in our disk-averaged temperature data. 
 
\begin{table}
\caption{Summary of observed and modeled millimeter disk-averaged brightness temperatures.}
\begin{tabular}{llllll}
\hline
Center Frequency (GHz) & Facility & UT Date & Observed T$^{a}_{\text{b}}$ (K) & Wet Disk Model T$^{b}_{\text{b}}$ (K) & Noise$^c$ (K) \\
\hline
\hline
3.096 & VLA & 02-Sep-2015 & $238.6\pm11.5$ & 223.8 & 1.3  \\
4.915 & VLA & 28-Jun-2003 & $215.1\pm10.8$ & 206.7 & --  \\
5.861 & VLA & 01-Sep-2015 & $214.3\pm10.9$ & 199.6 & 0.6  \\
8.328 & VLA & 27-Jun-2003 & $183.3\pm9.2$ & 184.2 & --  \\ 
9.869 & VLA & 01-Sep-2015 & $177.2\pm8.9$ & 177.7 & 0.4  \\ 
14.880 & VLA & 01-Sep-2015 & $153.3\pm7.7$ & 161.4 & 0.7  \\
14.990 & VLA & 26-Jun-2003 & $169.7\pm8.5$ & 161.1 & -- \\
23.061 & VLA & 11-Oct-2003 & $150.6\pm7.5$ & 148.6 & -- \\
32.958 & VLA & 01-Sep-2015 & $158.1\pm7.9$ & 141.7 & 2.0 \\
42.827 & VLA & 12-Oct-2003 & $147.4\pm7.4$ & 138.0 & -- \\
95.012 & ALMA & 07-Jul-2017 & $126.6\pm6.3$ & 121.6 & 0.1  \\ 
96.970 & ALMA & 07-Jul-2017 & $126.0\pm6.3$ & 121.0 & 0.1  \\ 
107.000 & ALMA & 07-Jul-2017 & $120.5\pm6.0$ & 117.9 & 0.2 \\ 
109.000 & ALMA & 07-Jul-2017 & $118.8\pm6.0$ & 116.7 & 0.3  \\
135.986 & ALMA & 08-Oct-2016 & $108.5\pm5.4$ & 105.8 & 0.3 \\ 
137.924 & ALMA & 08-Oct-2016 & $108.0\pm5.4$ & 105.2 & 0.2  \\  
147.986 & ALMA & 08-Oct-2016 & $104.3\pm5.2$ & 102.7 & 0.2  \\ 
149.986 & ALMA & 08-Oct-2016 & $104.5\pm5.2$ & 102.3 & 0.3  \\
223.982 & ALMA & 24-Oct-2016 & $93.4\pm4.7$ & 92.1 & 0.4  \\ 
225.982 & ALMA & 24-Oct-2016 & $93.0\pm4.7$ & 91.9 & 0.4  \\ 
239.981 & ALMA & 24-Oct-2016 & $93.1\pm4.7$ & 90.0 & 0.6  \\ 
241.981 & ALMA & 24-Oct-2016 & $92.8\pm4.6$ & 89.8 & 0.6 \\
\hline
\end{tabular}\label{table:vla-disk-avg}

\footnotesize{$^a$ The listed errors are the absolute errors, estimated at $5\%$ from the calibrators.} \\
\footnotesize{$^b$ The model brightness temperatures from the best-fitting wet adiabat MCMC retrievals.}\\
\footnotesize{$^c$ Random errors defined as the RMS on the sky RMS errors are not available for the VLA 2003 data}.

\end{table}


\section{Modeling}

We generate models of Neptune's brightness temperature using the radiative transfer (RT) code Radio-BEAR\footnote{This code is available at: \url{https://github.com/david-deboer/radiobear} .}. Radio-BEAR generates synthetic spectra by solving the equation of radiative transfer through a model atmosphere. For a fuller description of Radio-BEAR, we refer the reader to \citet{dePater2005}, \citet{dePater2014}, and \citet{dePater2019}, which outline how the model atmosphere is constructed from bottom-up and details the absorption coefficients and line profiles for the species in this work. A description of the temperature profiles, cloud structure, and compositions considered in the model atmosphere is given below.

\subsection{Temperature Profile}

Temperature profiles are calculated from deep in the atmosphere upward, assuming either a dry or wet adiabat such that the profile matches the \textit{Voyager} 2 temperature of 71.5 K at 1 bar \citep{Lindal1992}. At pressures shallower than 1 bar, the temperature profile follows that derived from mid-infrared inversions by \citet{Fletcher2014}. Temperatures, pressures, and altitudes are related through hydrostatic equilibrium. We assume that the atmosphere is in local thermodynamic equilibrium. 

\begin{sidewaysfigure}
\centering
  \includegraphics[width=\linewidth]{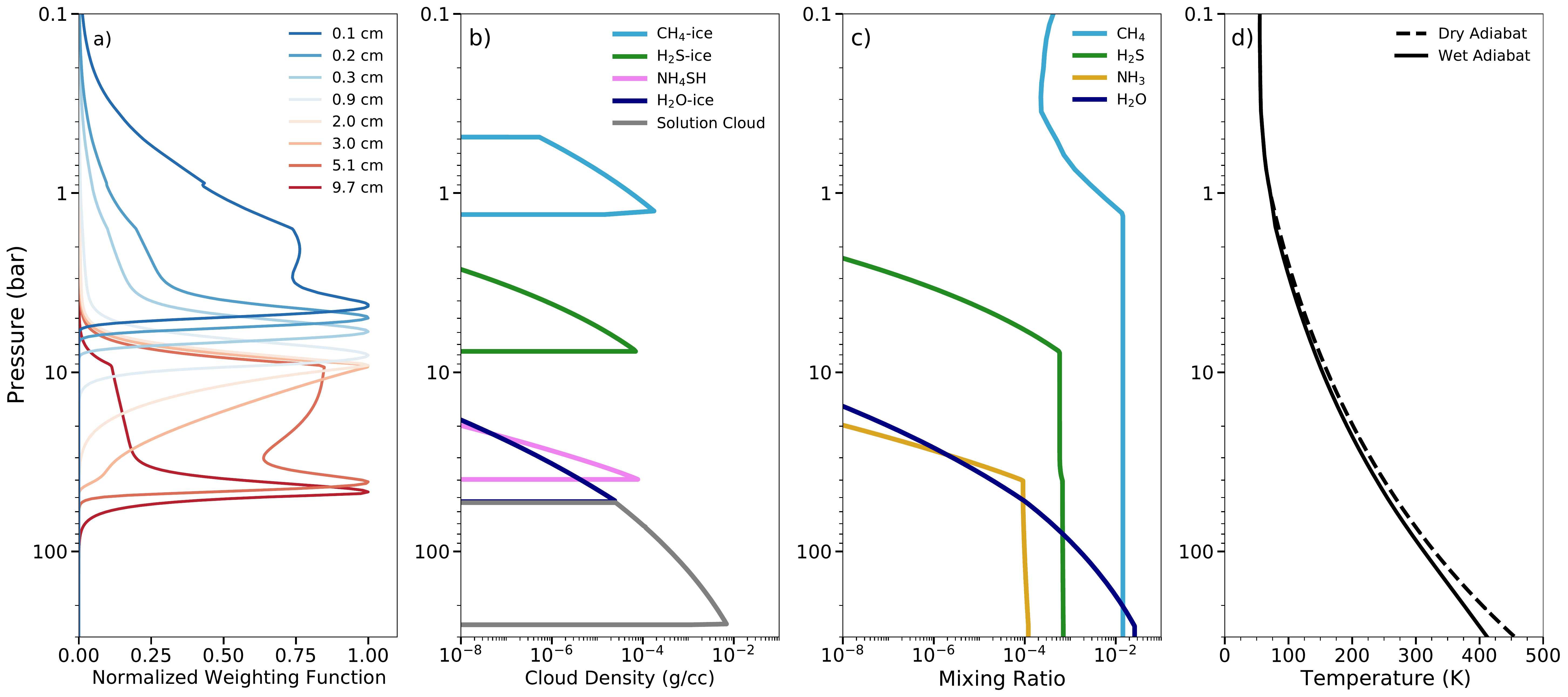}
\caption{The normalized weighting functions at nadir for the wavelengths used in this study (a), showing the pressure at which each observation is sensitive. Different gas profiles, temperatures, and geometries will alter the peaks of the weighting functions. The clouds expected to form on Neptune as a function of their density and condensation pressures are shown in (b). The cloud density and condensation pressure is determined from thermochemical equilibrium and the nominal gas profiles, pictured in (c), which assume 30$\times$ protosolar CH$_4$ (light blue), H$_2$S (green) and H$_2$O (dark blue), and $1\times$ protosolar NH$_3$ (yellow) at the deepest levels (below all cloud formation). The cloud and gas structure is also linked to the assumed temperature profile; both the dry and wet adiabatic temperature profiles for the nominal gas abundances are plotted in (d).} 
\label{fig:conts-clouds-gas-temp}
\end{sidewaysfigure}

Figure \ref{fig:conts-clouds-gas-temp} plots the wet and dry temperature profiles used in this paper. These are derived from a `nominal' atmosphere with 30$\times$ protosolar abundances of H$_2$S, CH$_4$, and H$_2$O, with 1$\times$ NH$_3$, no PH$_3$, `intermediate' H$_2$ (see: Sec. 3.5), and 100$\%$ relative humidity for each condensible species\footnote{We use the protosolar values from \citet{Asplund2009}: C/H$_2$ = 5.90E-4;
N/H$_2$ = 1.48E-4; O/H$_2$ = 1.07E-3; S/H$_2$ = 2.89E-5.}. Figure \ref{fig:conts-clouds-gas-temp} also plots the nominal abundances of Neptune's condensibles. 

\subsection{Cloud Structure}

The condensibles in Neptune's upper atmosphere are H$_2$S, CH$_4$, H$_2$O, and NH$_3$, all of which may condense to form clouds. Clouds expected to form on Neptune, from bottom-up, include: an aqueous ammonia solution (H$_2$O-NH$_3$-H$_2$S), water-ice, ammonium hydrosulfide (NH$_4$SH), H$_2$S- or NH$_3$-ice (whichever is left over after NH$_4$SH formation), and CH$_4$-ice \citep{Weidenschilling1973, Atreya2005}. The cloud density may affect microwave measurements. However, little is known about the cloud density on Neptune and clouds have been shown to not affect the microwave opacity on Jupiter \citep{dePater2019}. Therefore, we ignore the effect of cloud opacity in our modeling and focus instead on the effect of gas opacity. 

\subsection{Condensible Species}

The gas opacity of the microwave spectrum for gas giant atmospheres is dominated by H$_2$S, NH$_3$, H$_2$O, and the collision-induced absorption (CIA) of H$_2$ (we include: H$_2$-H$_2$, H$_2$-He, and H$_2$-CH$_4$) \citep{dePater1993}. To form the NH$_4$SH cloud, H$_2$S and NH$_3$ are reduced in equal molar quantities until the product of their partial pressures reaches the equilibrium constant of the reaction forming NH$_4$SH. On Uranus and Neptune, the observation that NH$_3$ is absent while H$_2$S is present above the NH$_4$SH layer \citep{dePater1991, Irwin2018, Irwin2019a, Molter2020} implies that this process takes up all of the NH$_3$, leaving an excess of H$_2$S gas above the layer. Therefore, while NH$_3$ and H$_2$O are strong microwave absorbers, they are only abundant deeper than $\sim40$ bar and only impact Neptune's radio spectrum significantly at wavelengths longer than 10 cm. This is shown in Figure \ref{fig:conts-clouds-gas-temp}, which plots the normalized weighting functions at each observed wavelength assuming a nominal atmosphere. Our ALMA and VLA observations, with wavelengths shorter than 10 cm, are sensitive to pressures between $1-50$ bar, peaking at altitudes at and above the aqueous NH$_3$, H$_2$O-ice, and NH$_4$SH cloud formations. While NH$_3$ can not be probed directly with our wavelength coverage, the chemical connection between H$_2$S and NH$_3$ means its abundance can be inferred. We therefore allow the NH$_3$ profile to vary unlike in \citet{Tollefson2019}. In our model, the formation of the NH$_4$SH cloud is governed by equilibrium chemistry described in \citet{Lewis1969}.

Our forward models allow the `deep' abundances of gaseous H$_2$S, NH$_3$, and CH$_4$ to vary.  We define `deep' as pressures below the NH$_4$SH cloud (forming at $\sim 50$ bar) but above aqueous solution cloud formation ($P > 100$ bar). For the nominal abundances, the solution cloud in full thermochemical equilibrium removes about 5$\%$ of the interior H$_2$S and 25$\%$ of the interior NH$_3$. From our retrieved values of `deep' H$_2$S and NH$_3$, we obtain abundances below all cloud formation. We set no \textit{a priori} restriction on the `deep' H$_2$S or NH$_3$ abundance, meaning either may survive above NH$_4$SH formation. H$_2$S or NH$_3$, whichever persists, will then form an ice cloud and we allow its relative humidity to vary (see Fig. \ref{fig:conts-clouds-gas-temp}, which shows H$_2$S-ice forming, as the `deep' H$_2$S abundance is larger than that of NH$_3$). 

\subsubsection{Phosphine (PH$_3$)}

PH$_3$ is an important disequilibrium species on giant planets, tracing both chemistry and convective motion. In the deep atmosphere, PH$_3$ should oxidize to form P$_4$O$_6$ and dissolve in water \citep{Fegley1986}. In the upper atmosphere, PH$_3$ is photolyzed and subsequent photochemical reactions may form to produce P$_4$ or complex polymers an compounds. Thus, PH$_3$ must be rapidly uplifted from the deep atmosphere to exist, making it useful as a passive tracer for both horizontal and vertical motions \citep{Fegley1986, Fletcher2009}.  

PH$_3$ has not been detected on Uranus or Neptune, but if present is an important microwave absorber \citep{DeBoer1996}. The effect is most prominent at the PH$_3$ ($1-0$) rotation line at 266.9 GHz. The radio observations analyzed here only reach frequencies as high as 242 GHz (ALMA Band 6), meaning only the wings of the absorption line are detectable. However, if the abundance of PH$_3$ is large enough, the pressure-broadened rotation line will be detectable in our highest frequency data. Thus, upper limits on the uplifted PH$_3$ abundance can be placed. 

Our forward model assumes a constant PH$_3$ abundance until the temperature and pressure become so low that PH$_3$ condenses ($\sim$ 1 bar). We also only include PH$_3$ in our retrievals at latitudes where we expect upwelling and enriched condensibles, i.e. latitudes with comparatively cold brightness temperatures. 

\subsection{\textit{ortho/para} H$_2$}

The ortho-/para-H$_2$ fraction also influences Neptune's radio brightness temperature by modifying both the adiabatic lapse rate and the gas opacity \citep{Trafton1967, Wallace1980, dePater1985, dePater1993}. The ratio of ortho- to para-hydrogen in equilibrium depends on temperature; however, fast vertical mixing could bring the ratio of ortho and para states of hydrogen away from equilibrium and toward a ``normal" ratio of three parts ortho to one part para. At latitudes where fast vertical mixing is unlikely, H$_2$ is presumed to exist in an ``intermediate" state, proposed by \citet{Trafton1967}. In this case, the ortho and para states (which define the CIA properties) are set to the equilibrium value at the local temperature while the specific heat is set near that of ``normal" hydrogen. ``Intermediate" hydrogen is discussed further in \citet{Massie1982}, who provide physical reasons for the choice of this ``intermediate” state. ``Normal" hydrogen has been shown to decrease the microwave brightness temperature relative to ``intermediate" hydrogen \citep{dePater1993, LuszczCook2013, Tollefson2019}. 

In our retrievals, we parameterize the state of H$_2$ as between 0.0 and 1.0, where 0.0 represents the fully ``normal'' state while 1.0 is fully ``intermediate." Values in between represent a weighted average of the absorption coefficient between the two states. The specific heat is always set near to that of ``normal'' hydrogen, regardless of the parameter state.

\begin{figure}
\centering
  \includegraphics[width=\linewidth]{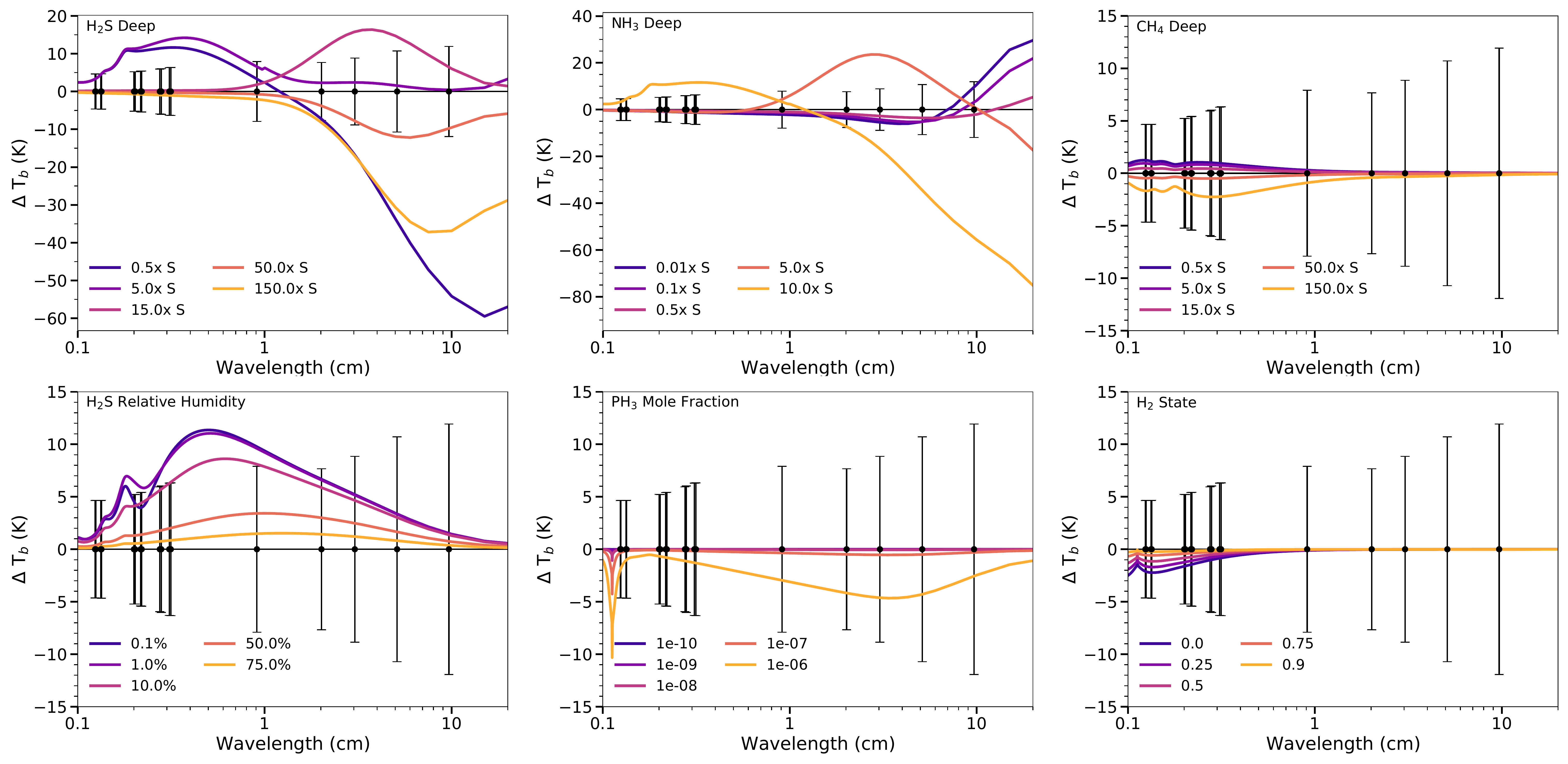}
\caption{The effect of changing modeled parameters on Neptune's disk-averaged microwave spectrum following a wet adiabat. Each panel shows how varying only the listed parameter from the nominal atmosphere alters the disk-averaged spectrum. $\Delta T_b$ is the difference from the nominal spectrum, with colored lines representing different values of the varied parameter. Positive $\Delta T_b$ values mean that the new model is warmer than the nominal model. The deep H$_2$S, NH$_3$, and CH$_4$ abundances are given in terms of their protosolar enhancements (top row). The relative humidity of H$_2$S at its ice-cloud formation, the mole fraction of PH$_3$, and the H$_2$ state (defined in section 3.4) are also  (bottom row). Interesting behavior is seen in the deep H$_2$S panel, as both large and small amounts of H$_2$S (0.5$\times$ protosolar, blue vs. 150$\times$ protosolar, yellow) produce negative $\Delta T_b$ longward of 1 cm, while only depleted amounts of H$_2$S result in positive $\Delta T_b$ in the millimeter, relative to the nominal model. These trends are explained in Section 3.6. The central wavelengths of our maps are plotted with black points. The error bars represent the calibration errors to give a sense of how important each parameter is in fitting the model to the data.}
\label{fig:abu-vs-spec}
\end{figure}

\subsection{Retrievals}

In order to estimate uncertainties in our model parameters, we couple our Radio-BEAR models to Markov Chain Monte Carlo (MCMC) simulations via a python implementation of the \citet{Goodman2010} ensemble sampler called \textit{emcee} \citep{Foreman2013}. \textit{emcee} has been used in near-infrared analyses of Uranus' and Neptune's hazes \citep{deKleer2015,LuszczCook2016}. We use similar log-likelihood Gaussian function and uniform/log-uniform priors as these authors; letting $\theta$ represent the free parameters in the model, the likelihood function $\ln p$ is:

\begin{equation}
    \text{ln } p(T_b | \nu, \theta, \sigma) = -0.5 \sum_{\nu_n} \left[\left(\frac{T_{b,n} - T_{b,m}}{\sigma_n}\right)^2 + \text{ln} \left(2 \pi \sigma^2_n\right)\right],
\end{equation}

where $T_{b,n}$ is the observed brightness temperature, $T_{b,m}$ is the RT modeled brightness temperature, and $\sigma_n$ is the total uncertainty, all at a given wavelength $\nu$. This approach is in contrast to the methods in \citet{Tollefson2019} who compared the ALMA data to forward models of Neptune's atmosphere. From these forward models, they obtained deep abundances of H$_2$S and CH$_4$ that fit the latitudinally-varying brightness temperatures. A downside to this approach is that one can only rule out models which are improbable from $\chi^2$-statistics, meaning uncertainties are not retrieved.

We use 30 walkers and let MCMC run for between $4000-5000$ steps. This run length ensured that the autocorrelation time was sufficiently shorter than the number of steps. All plots and tables show the distributions after burn-in, where the range of retrieved values approximately appear like their final probability density. The burn-in phase ends after 50$\%$ of the steps have been completed.

Observed and modeled brightness temperatures are obtained in each of the identified latitude bins listed in Section 2.2. One issue with RT models of particular regions on Neptune is that the finite size of the point-spread function (PSF) results in blurring of the disk. That is, the temperature within a particular latitude bin is a convolution of that latitude region, nearby latitudes, and sometimes the background sky. This effect is hard to model while conducting MCMC for two reasons: convolving the whole disk with the PSF is computationally expensive; and the model composition of the surrounding latitudes must be simultaneously known in order to obtain the exact temperature distribution across the disk. We circumvent this issue in two steps. First, we generate limb-darkened model disks of the best-fitting disk-averaged models (see Section 4.1) and then determine the brightness temperature in the region of interest with and without convolving the disk with the PSF. The ratio of the PSF convolved disk-averaged model temperature to that without convolution is called the PSF-scale. We multiply each MCMC retrieved model by the PSF-scale to obtain our final model brightness temperature incorporating the effect of convolution. This model temperature is fed into the \textit{emcee} likelihood function. The effect of this scaling is smallest near the center of the disk and largest near the limb. Second, we add an error term to account for the uncertainty introduced by this approach, 

\begin{equation}
\sigma_{T}^2 = \sigma_{CAL}^2 + \sigma_{PSF}^2, 
\end{equation}

where $\sigma_{T}$ is the total error term used by \textit{emcee} to calculate the likelihood function, $\sigma_{CAL}$ is the $5\%$ calibration error and $\sigma_{PSF}$ is the error introduced by RT-modeling not accounting for the PSF. In our MCMC retrievals, we define an additional free parameter, $\sigma$, that is proportional to $\sigma_{PSF}$:

\begin{equation}
\sigma_{PSF} = \sigma \cdot T \cdot S,
\end{equation}

where $T$ is the observed brightness temperature and $S$ is the product of the beam's semi-major and semi-minor axes. Adding an extra free parameter in the uncertainty is general practice in MCMC, as it encompasses unknown sources of error (for instance, via the introduction of PSF-scale). We find that the maximum retrieved values of $\sigma_{PSF}$ are never larger than $5\%$ of the product of $T$ and $S$ at disk center and $15\%$ at the limb.

\subsection{Atmospheric Models and Free Parameters}

We consider four types of atmospheric models for these retrievals, depending on the latitude bin and expected dynamics, described below. Table \ref{table:atm+params} lists each of these models and the allowed free parameters. Figure \ref{fig:abu-vs-spec} summarizes the effect of altering these parameters on the disk-averaged microwave spectrum of Neptune relative to the nominal model. Of note, the first two panels highlight the strong interaction between the deep H$_2$S and NH$_3$ abundances as a function of wavelength. NH$_3$ can exceed H$_2$S, resulting in NH$_3$ surviving NH$_4$SH cloud formation, by either sufficiently depleting H$_2$S (blue line, 0.5$\times$ S, in the H$_2$S deep panel) or enriching NH$_3$ (yellow line, 10$\times$ S, in the NH$_3$ deep panel). The millimeter spectrum becomes brighter, as more NH$_3$ will reacts with more H$_2$S at NH$_4$SH formation, thereby reducing the total amount of H$_2$S above it. On the other hand, the centimeter spectrum becomes colder, as NH$_3$ gas is a stronger microwave absorber than H$_2$S gas. All other parameters have the strongest impact at $\lambda \leq 1$ cm. Note, however, that our ability to retrieve the deep CH$_4$ abundance is limited compared to that of H$_2$S. This is due to a combination of having less overall impact on the microwave opacity, lower than our calibration errors, and other sources like the H$_2$S relative humidity, PH$_3$, and the H$_2$ state having comparable or stronger effects in the mm. The considered models are:

\begin{enumerate}
    \item Enriched Atmosphere: Used for latitude bands which have relatively cold brightness temperatures, where enriched and upwelling air is expected. The `deep'\footnote{Again, `deep' is defined as pressures above the H$_2$O and aqueous solution clouds and below the NH$_4$SH cloud.} abundances of of H$_2$S, NH$_3$, and CH$_4$ are allowed to vary. The NH$_4$SH, H$_2$S-ice and CH$_4$-ice clouds are allowed to form and the relative humidity of H$_2$S is varied. The PH$_3$ abundance is allowed to vary; a uniform vertical profile is assumed up to the saturation pressure of PH$_3$. The ortho- to para- H$_2$ fraction is allowed to vary between $0.0-1.0$, representing the range of fully `normal' to fully `intermediate' states. Both dry and wet adiabats are considered.

    \item Depleted Atmosphere: Used for latitude bands which have relatively warm brightness temperatures, where depleted and downwelling air is expected. The `deep' abundance of condensibles is set to that of the enriched atmosphere at altitudes below the `mixing pressure:' $P_{\text{mix}}$. At shallower altitudes, the abundance of condensibles is varied by some fraction of the deep abundance. The final profiles for the condensibles look like step functions, with the transition at $P_{\text{mix}}$. We allow the formation of H$_2$S- and NH$_3$-ice. Both dry and wet adiabats are considered.

    \item Disk Average: The disk-averaged model for the atmosphere looks like that of the enriched, but we only let the `deep' abundances of H$_2$S, NH$_3$, and CH$_4$, and the relative humidity of H$_2$S to vary. This model is done to compare to prior work and so both PH$_3$ and the H$_2$ state are not varied. We caution that disk-averaged retrievals poorly describe the physics within the planet. Both dry and wet adiabats are considered. 
    
    \item Standard: At other latitudes not fitting the above prescriptions, we model the atmosphere similar to the enriched atmosphere as it follows standard cloud formation and thermochemical equilibrium unlike the depleted/downwelling atmosphere. Unlike the `Enriched' model, the PH$_3$ abundance and ortho/para H$_2$ state are not allowed to vary. Both dry and wet adiabats are considered.

\begin{sidewaystable}
\centering
\caption{Summary of atmospheric models and free parameters used at each latitude band. Parameters marked with an $\times$ are varied in MCMC. If they are not varied, their set value is given instead, if applicable. In our retrievals, all free parameters are varied in log$_{10}$-space apart from the H$_2$ state.}

\begin{tabular}{llcccccccccccc}
\hline
 Latitude Band & Model & H$_2$S Deep & NH$_3$ Deep & CH$_4$ Deep & PH$_3$ & H$_2$ State$^a$ & H$_2$S RH$^b$ & H$_2$S Frac$^c$ & NH$_3$ Frac$^c$ & CH$_4$ Frac$^c$ & $P_{\text{mix}}$ & $\sigma_{\text{PSF}}$ \\
 \hline
\hline
4$^{\circ}$N $-$ 20$^{\circ}$N & Standard & $\times$ & $\times$ & $\times$ & 0.0 & Intermediate & $\times$ &  N/A &  N/A &  N/A & N/A & $\times$ \\
12$^{\circ}$S $-$ 4$^{\circ}$N & Standard & $\times$ & $\times$ & $\times$ & 0.0 & Intermediate & $\times$ &  N/A &  N/A &  N/A & N/A & $\times$\\
36$^{\circ}$S $-$ 12$^{\circ}$S & Enriched & $\times$ & $\times$ & $\times$ & $\times$ & $\times$ & $\times$ &  N/A &  N/A &  N/A & N/A & $\times$\\
50$^{\circ}$S $-$ 36$^{\circ}$S & Standard & $\times$ & $\times$ & $\times$ & 0.0 & Intermediate & $\times$ &  N/A &  N/A &  N/A & N/A & $\times$ \\
66$^{\circ}$S $-$ 50$^{\circ}$S & Standard & $\times$ & $\times$ & $\times$ & 0.0 & Intermediate & $\times$ &  N/A &  N/A &  N/A & N/A & $\times$ \\
90$^{\circ}$S $-$ 66$^{\circ}$S & Depleted & $\times$ & $\times$ & $\times$ & 0.0 & Intermediate & $\times$ & $\times$ & $\times$ & $\times$ & $\times$ & $\times$ \\
Global & Disk-Average & $\times$ & $\times$ & $\times$ & 0.0 & Intermediate & $\times$ &  N/A &  N/A &  N/A & N/A & Absent \\
 \hline
\end{tabular}\label{table:atm+params}
\raggedright
\footnotesize{$^a$ All retrieved parameters in this paper except H$_2$ State are varied in log$_{10}$ space. } \\
\footnotesize{$^b$ RH stands for relative humidity. Supersaturated H$_2$S is allowed in Section 4.3.} \\
\footnotesize{$^c$ Frac is the fractional abundance relative to the 'deep' amount of the condensibles existing above the mixing pressure $P_{\text{mix}}$ in the 'Depleted' south polar model.} 

\end{sidewaystable}

\end{enumerate}

\section{Results}

\subsection{Disk-Averaged Profiles}

We calculate Neptune's disk-average brightness temperature by totaling the flux density contained within Neptune's disk plus the area extending out three times the model beam diameter past the limb. The apparent brightness temperature measured in interferometric imaging is less than the true value by an amount related to the cosmic microwave background (CMB). The CMB correction is applied following the procedure laid out in Appendix A of \citet{dePater2014}. These results are combined with measurements of Neptune's disk-average brightness temperature from VLA 2003 \citep{dePater2014} and ALMA \citep{Tollefson2019} to form our set of data used in MCMC retrievals. A summary of this data set is given in Table \ref{table:vla-disk-avg}.

\begin{figure}
\centering
  \includegraphics[width=\linewidth]{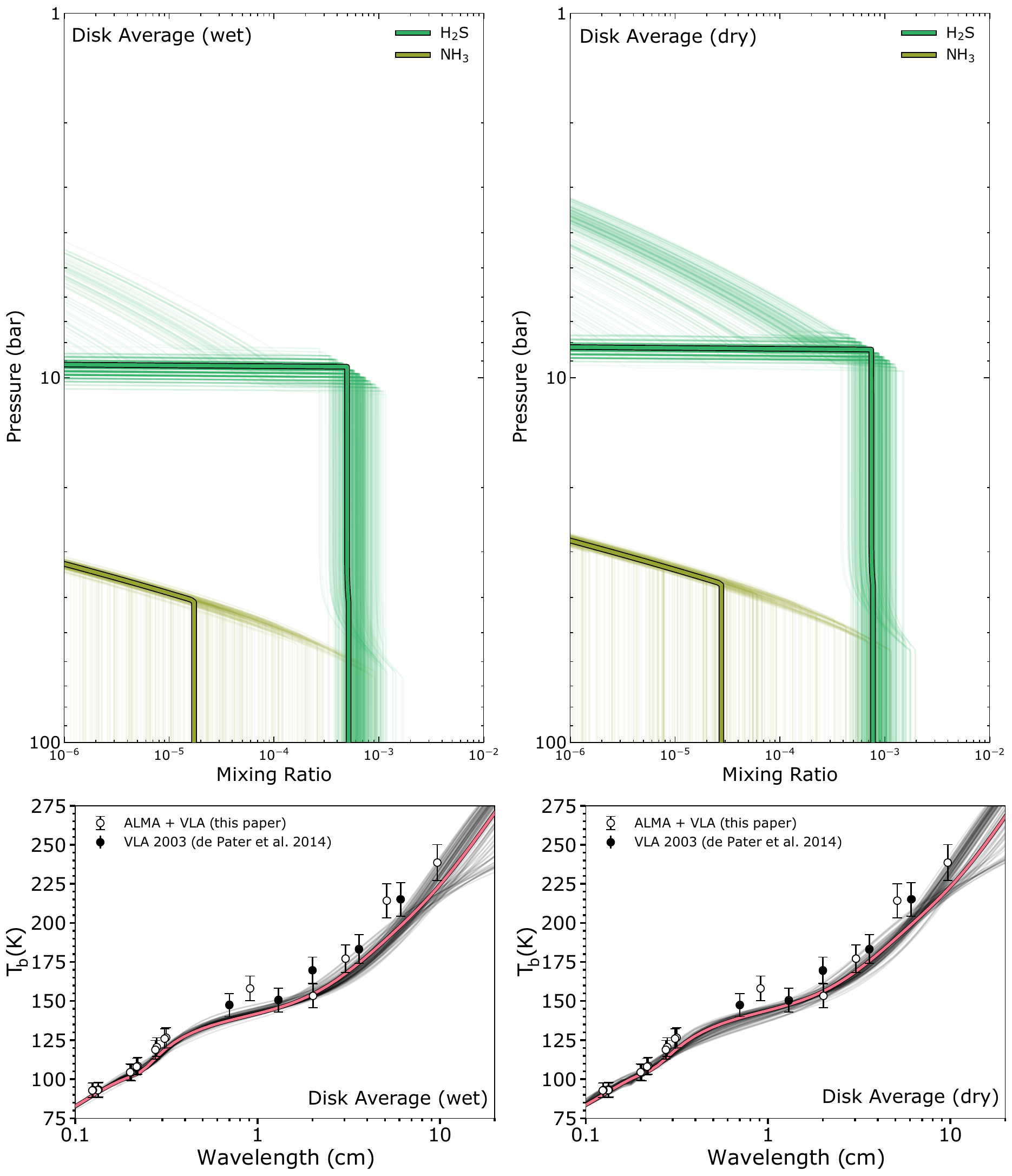}
\caption{Top row: Best fitting H$_2$S (thick green) and NH$_3$ (thick yellow) profiles for the disk-average wet (left) and dry (right) adiabat models. Thin lines are 500 random retrieved profiles. Bottom row: Disk-averaged data from VLA 2003 (de Pater et al. 2014; black circles) and from ALMA and the upgraded VLA (this paper, white circles). The thick red line is the best fitting spectrum following a wet (left) and dry (right) adiabat. Thin black lines are also plotted, representing 100 random draws from the posterior.}
\label{fig:disk-avg-abu-spec}
\end{figure}

Figure \ref{fig:disk-avg-abu-spec} plots the best-fitting H$_2$S and NH$_3$ profiles and 500 random retrieved profiles from the posterior distribution for a wet adiabat. Figure \ref{fig:disk-avg-abu-spec} also shows the observed disk-averaged brightness temperatures versus wavelength, likewise plotting the best-fitting and 100 random retrieved model spectra assuming a wet adiabat. Table \ref{table:disk-avg-results} lists the retrieved parameters. H$_2$S is more abundant than NH$_3$ in both the dry and wet thermal profiles. There is also significant positive correlation between the deep NH$_3$ and H$_2$S abundances. This is expected based on the NH$_4$SH cloud chemistry; retrievals with larger NH$_3$ require additional H$_2$S to remove the NH$_3$ during the NH$_4$SH reaction to match the observations probing above the NH$_4$SH cloud. The dry adiabat permits marginally larger abundances of all condensibles throughout the upper atmosphere due to the need to increase the opacity in order to offset the higher temperatures.

\begin{table}
\centering
\begin{tabular}{|c|c|c|}
\hline
\multicolumn{3}{ |c| }{Disk Average, Wet Adiabat} \\
\hline
Parameter & Retrieved values & Protosolar Enhancement ($\times$ Solar) \\
\hline
H$_2$S below NH$_4$SH & $6.3^{+1.9}_{-1.2}\times10^{-4}$ & $26.8^{+8.1}_{-5.1}$\\
NH$_3$ below NH$_4$SH & $\leq 6.5\times10^{-4}$ & $\leq 5.4$\\
CH$_4$ & $\leq 1.6\times10^{-2}$ & $\leq 33.3$ \\
H$_2$S $H_{\text{rel}}$ & $\leq 18\%$ & --- \\
\hline
H$_2$S below all clouds & $6.6^{+2.0}_{-1.3}\times10^{-4}$ & $28.2^{+8.5}_{-5.4}$ \\
\hline
\hline
\multicolumn{3}{ |c| }{Disk Average, Dry Adiabat} \\
\hline
Parameter & Retrieved values & Protosolar Enhancement \\
\hline
H$_2$S below NH$_4$SH & $8.8^{+2.0}_{-2.1}\times10^{-4}$ & $37.5^{+9.4}_{-8.9}$ \\
NH$_3$ below NH$_4$SH & Unconstrained & Unconstrained \\
CH$_4$& $\leq 1.3\times10^{-2}$ & $\leq 27.1$\\
H$_2$S $H_{\text{rel}}$ & $\leq 75\%$ & --- \\
\hline
H$_2$S below all clouds & $9.3^{+2.1}_{-2.2}\times10^{-4}$ & $39.4^{+9.9}_{-9.4}$ \\
\hline
\end{tabular}
\caption{MCMC fit results for the disk-average models. The listed H$_2$S abundances below NH$_4$SH cloud formation are the 50th percentile of the posterior distributions with the error bars corresponding to the 16th and 84th percentiles, i.e. the $1\sigma$ uncertainties. NH$_3$, CH$_4$, and the relative humidity of H$_2$S are not well-constrained in either model and so the 97.5th percentile value, representing the 2$\sigma$ upper limit, is given instead. Abundances below all cloud formation are estimated according to the discussion in Section 3.3}
\label{table:disk-avg-results}
\end{table}

\subsection{36$^{\circ}$S $-$ 12$^{\circ}$S and 90$^{\circ}$S $-$ 66$^{\circ}$S Profiles}

The latitude band between $36^{\circ}$S$-12^{\circ}$S is dark in both the VLA maps presented here (Fig. \ref{fig:vla-maps}) and ALMA residual maps. Low brightness temperatures imply increased opacity from Neptune's condensibles. Conversely, Neptune's south polar cap, between $90^{\circ}$S$-66^{\circ}$S, is warm in both datasets, implying a lower abundance of condensibles. Due to the lower abundance of condensibles, the peaks of the normalized weighting functions are deeper in the atmosphere than for the nominal model (Fig. \ref{fig:conts-clouds-gas-temp}), meaning we are more sensitive to abundances of Neptune's condensibles below the NH$_4$SH cloud at the south pole. As a result, we first obtain retrievals for the mixing pressure $P_{\text{mix}}$ and abundances of condensibles over the south polar cap. The retrieved probability density functions for the condensible abundances below $P_{\text{mix}}$ are then used as priors for the condensible abundance below the NH$_4$SH cloud in the cold mid-latitude band between $36^{\circ}$S$-12^{\circ}$S. The atmospheric models and free parameters used for these bands are described in Table \ref{table:atm+params} and Section 3.6.

The brightness temperatures fed into MCMC are obtained by averaging over all pixels that are within $\pm60$ degrees of the sub-observer longitude at each latitude band. Likewise, the modeled spectra are obtained using the average emission angle within this zonal average. The uncertainty is dominated by the flux calibration error. $\sigma_{\text{PSF}}$ is also added to this uncertainty, though its effect is small compared to the calibration error.

Data for Neptune's latitudinal variations are not available in the VLA 2003 maps apart from those at the south polar cap. We do not use these data as the temperatures reported in \citet{dePater2014} are the maximum temperatures within the south polar cap at each wavelength instead of the average.

\begin{table}
\centering
\begin{tabular}{|c|c|c|}
\hline
\multicolumn{3}{ |c| }{$90^{\circ}$S$-66^{\circ}$S, Wet Adiabat} \\
\hline
Parameter & Retrieved values & Protosolar Enhancement ($\times$ Solar) \\
\hline
H$_2$S below $P_{\text{mix}}$ & $0.3^{+15.7}_{-0.3}\times10^{-4}$ & $1.3^{+65.6}_{-1.3}$\\
NH$_3$ below $P_{\text{mix}}$ & $2.5^{+2.1}_{-2.1}\times10^{-4}$ & $2.1^{+1.7}_{-1.8}$\\
H$_2$S above $P_{\text{mix}}$ & $2.7^{+9.3}_{-2.7}\times10^{-7}$ & $1.2^{+3.9}_{-1.2}\times10^{-2}$ \\
NH$_3$ above $P_{\text{mix}}$ & $3.3^{+2.8}_{-2.2}\times10^{-7}$ & $2.8^{+2.3}_{-1.9}\times10^{-3}$\\
$P_{\text{mix}}$ &  $41^{+9}_{-7}$ bar & --- \\
\hline
\hline
\multicolumn{3}{ |c| }{$36^{\circ}$S$-12^{\circ}$S, Wet Adiabat} \\
\hline
Parameter & Retrieved values & Protosolar Enhancement ($\times$ Solar) \\
\hline
H$_2$S below NH$_4$SH & $8.3^{+2.5}_{-2.3}\times10^{-4}$ & $35.4^{+14.9}_{-9.8}$\\
NH$_3$ below NH$_4$SH & $1.9^{+2.0}_{-1.7}\times10^{-4}$ & $1.6^{+1.7}_{-1.4}$\\
CH$_4$ & $\leq 4.1\times10^{-2}$ & $\leq 85.4$ \\
H$_2$S $H_{\text{rel}}$ & $\leq 57\%$ & --- \\
PH$_3$ & $\leq8.2\times10^{-7}$ & $\leq 0.2$ \\ 
H$_2$ State & Unconstrained & --- \\
\hline
H$_2$S below all clouds & $8.7^{+2.6}_{-1.7}\times10^{-4}$ & $37.3^{+15.7}_{-10.3}$ \\
NH$_3$ below all clouds & $2.5^{+2.7}_{-2.3}\times10^{-4}$ & $2.1^{+2.3}_{-1.9}$ \\
\hline
\end{tabular}
\caption{MCMC results for the simultaneous fit to the $36^{\circ}$S$-12^{\circ}$S and $90^{\circ}$S$-66^{\circ}$S bands assuming a wet adiabat. The deep abundances of H$_2$S, NH$_3$, and CH$_4$, and $\sigma_{\text{PSF}}$ are the same in each region. Otherwise, parameters are varied according to Table \ref{table:atm+params}. The listed values are the 50th percentile of the posterior distributions with the error bars corresponding to the 16th and 84th percentiles, i.e.  the $1\sigma$ uncertainties. Parameters which are poorly constrained have their 97.5th percentile value listed instead, representing the $2\sigma$ upper limit. Abundances below all cloud formation are estimated according to the discussion in Section 3.3}
\label{table:lat-sim-wet-results}
\end{table}

\begin{table}
\centering
\begin{tabular}{|c|c|c|}
\hline
\multicolumn{3}{ |c| }{$90^{\circ}$S$-66^{\circ}$S, Dry Adiabat} \\
\hline
Parameter & Retrieved values & Protosolar Enhancement ($\times$ Solar) \\
\hline
H$_2$S below $P_{\text{mix}}$ & $3.6^{+25.4}_{-3.3}\times10^{-4}$ & $15.3^{+108.3}_{-8.9}$\\
NH$_3$ below $P_{\text{mix}}$ & $3.8^{+2.2}_{-2.4}\times10^{-4}$ & $3.2^{+1.8}_{-2.1}$\\
H$_2$S above $P_{\text{mix}}$ & $7.1^{+5.9}_{-2.8}\times10^{-6}$ & $3.0^{+2.5}_{-1.0}\times10^{-1}$ \\
NH$_3$ above $P_{\text{mix}}$ & $1.6^{+3.1}_{-1.3}\times10^{-7}$ & $1.3^{+2.6}_{-1.0}\times10^{-3}$\\
$P_{\text{mix}}$ &  $33^{+6}_{-4}$ bar & --- \\
\hline
\hline
\multicolumn{3}{ |c| }{$36^{\circ}$S$-12^{\circ}$S, Dry Adiabat} \\
\hline
Parameter & Retrieved values & Protosolar Enhancement ($\times$ Solar) \\
\hline
H$_2$S below NH$_4$SH & $1.2^{+0.4}_{-0.3}\times10^{-3}$ & $51.1^{+17.1}_{-12.7}$\\
NH$_3$ below NH$_4$SH & $3.5^{+1.9}_{-2.8}\times10^{-4}$ & $2.9^{+1.6}_{-2.3}$\\
CH$_4$ & Unconstrained & --- \\
H$_2$S $H_{\text{rel}}$ & $46^{+31}_{-34}\%$ & --- \\
PH$_3$ & $\leq4.4\times10^{-7}$ & $\leq0.1$ \\ 
H$_2$ State & Unconstrained & --- \\
\hline
H$_2$S below all clouds & $1.3^{+0.4}_{-0.3}\times10^{-3}$ & $53.8^{+18.0}_{-13.4}$ \\
NH$_3$ below all clouds & $4.7^{+2.5}_{-3.7}\times10^{-4}$ & $3.9^{+2.1}_{-3.1}$ \\
\hline
\end{tabular}
\caption{As Table \ref{table:lat-sim-wet-results}, but assuming a dry adiabat in each region.}
\label{table:lat-sim-dry-results}
\end{table}

\begin{figure}
\centering
  \includegraphics[width=\linewidth]{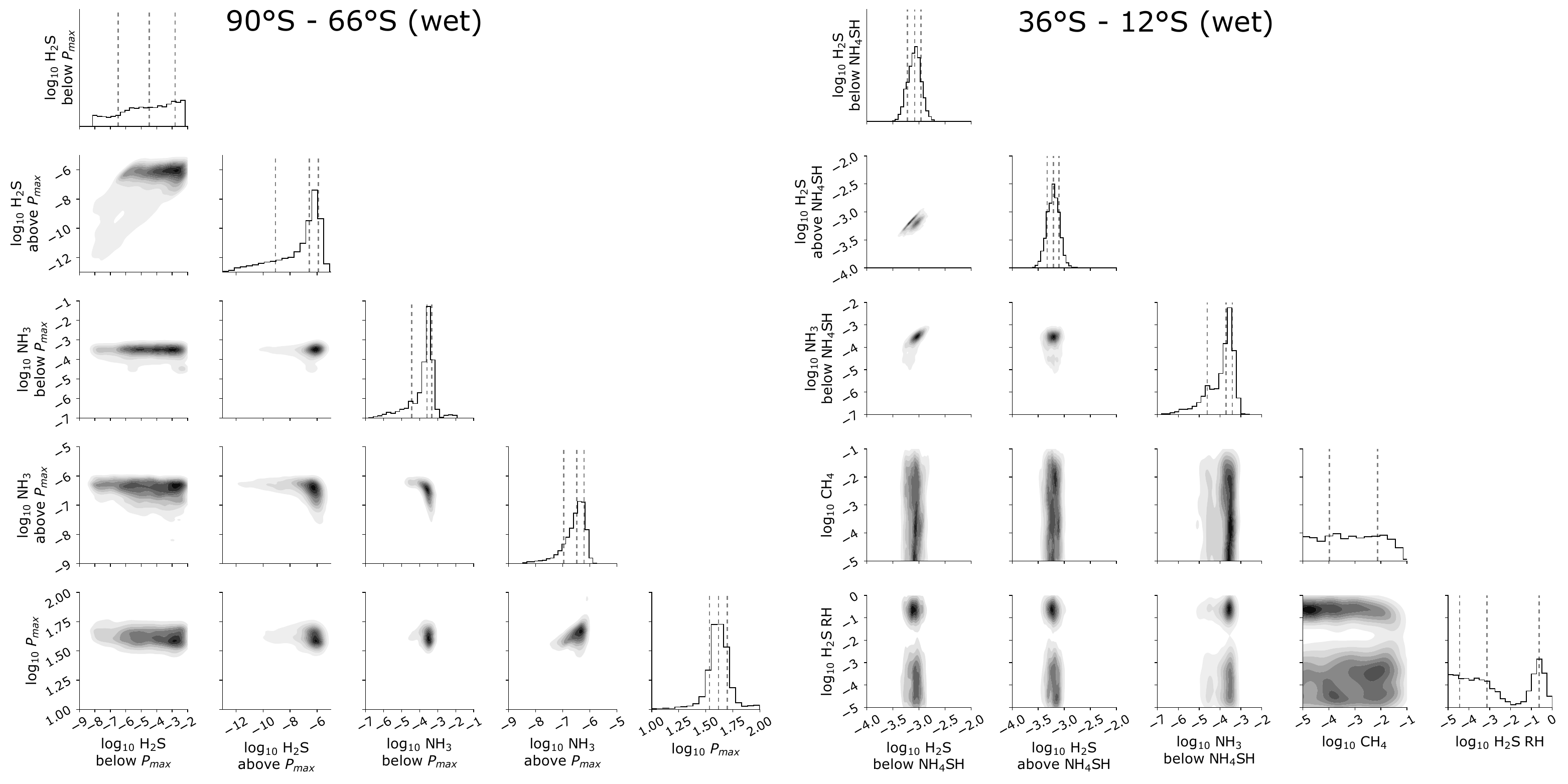}
\caption{Cornerplots of highlighted parameters that control the abundance profiles of the condensibles. Histograms of the retrieved distributions are plotted along diagonals, while the covariance between pairs of parameters are plotted on the off-diagonals. Darker regions on the off-diagonal plots indicate a larger density of retrieved solutions. The $90^{\circ}$S$-66^{\circ}$S (left) and $36^{\circ}$S$-12^{\circ}$S (right) latitude regions are shown, which both assume a wet adiabat. Dashed vertical lines mark the median and $1\sigma$ bounds.}
\label{fig:sim-corner}
\end{figure}

Tables \ref{table:lat-sim-wet-results} and \ref{table:lat-sim-dry-results} list the 16th/50th/84th percentiles on the gas profile parameters for the simultaneous $36^{\circ}$S$-12^{\circ}$S and $90^{\circ}$S$-66^{\circ}$S models. If a parameter is not well constrained, the 97.5th-percentile (2$\sigma$) upper limit is given instead. Corner plots showing the covariance and probability distribution for the free parameters are given in Figure \ref{fig:sim-corner}, assuming the temperature profile at the south polar cap follows a wet adiabat. As in the disk-average results, there is significant correlation between the deep H$_2$S and NH$_3$ abundances. The presence of additional NH$_3$ at deep levels will lead to additional H$_2$S at the NH$_4$SH layer. However, H$_2$S is the primary absorber above the NH$_4$SH layer and so our data are very sensitive to its abundance. Therefore an increased NH$_3$ abundance requires a compensating increase in the H$_2$S abundance. In addition, our results show that the H$_2$S profile is constrained at the cold mid-latitudes but unconstrained at the south pole. The NH$_3$ abundance below the NH$_4$SH cloud is somewhat constrained at each latitude band, though a long-tail forms at low abundances for retrievals with less H$_2$S.  In addition, $P_{\text{mix}}$ is well-constrained at the south polar cap and is inline with the NH$_4$SH condensation pressure. The retrievals also indicate a constant amount of NH$_3$ above $P_{\text{mix}}$ at the south polar cap, though the median values are larger than the 12 ppb abundance required by \citet{dePater2014} to best match their own observations.

The deep CH$_4$ abundance is only weakly constrained as its effect on the radio opacity is small compared to the calibration uncertainty (see Fig. \ref{fig:abu-vs-spec}). Moreover, it competes with effects from PH$_3$, the H$_2$ state, and the relative humidity of H$_2$S. At $36^{\circ}$S$-12^{\circ}$S, we obtain an upper limit of $85.4\times$ protosolar or $4.1\%$ mixing ratio at the 2$\sigma$ level, assuming a wet adiabat globally. These upper limits are consistent with the disk-average amount of CH$_4$ reported by \citet{Baines1995} at 2.2$\%$. However, the $4\%$ deep CH$_4$ abundance favored by \citet{Karkoschka2011} is similar to our 2$\sigma$ upper limit. A global dry adiaba permits larger abundances of each condensible, with the CH$_4$ abundance completely unconstrained. Our CH$_4$ results may be reconciled with differences in the assumed thermal profiles, as the thermal profile assumed by \citet{Karkoschka2011} is slightly warmer than a wet adiabat.

Our results are consistent with no PH$_3$ with upper limits of $\sim0.4-0.8$ ppm, depending on the assumed temperature model. We are completely insensitive to the ortho/para H$_2$ state. 

The relative humidity of H$_2$S is also unconstrained, apart from $2\sigma$ upper limits. As shown in Figure \ref{fig:abu-vs-spec}, the difference between relative humidities of a few percent or less on the retrieved spectra are minimal when compared to our uncertainties; depleting more and more H$_2$S has an inconsequential effect on the opacity once enough is removed. 

\begin{figure}
\centering
  \includegraphics[width=0.75\linewidth]{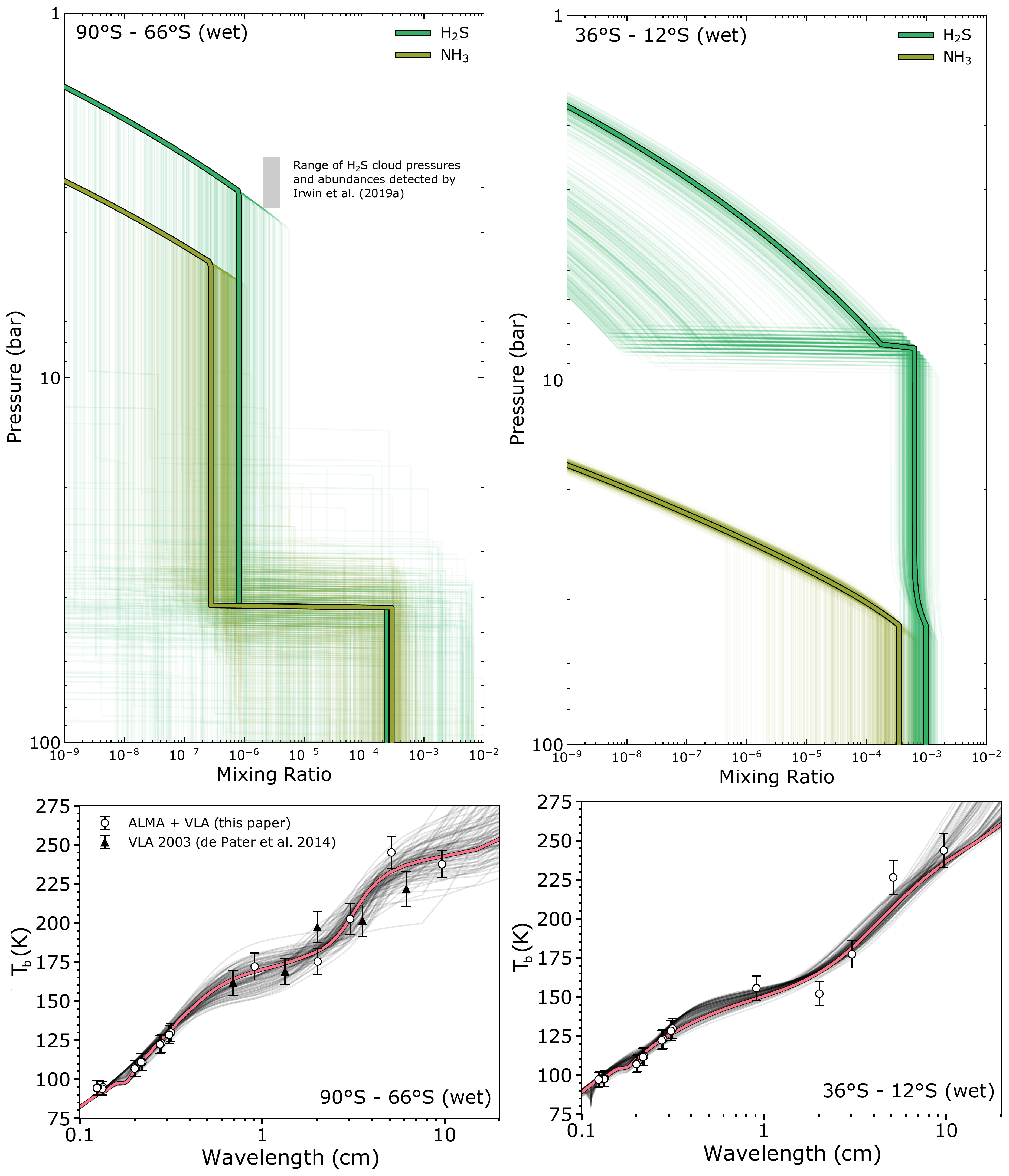}
\caption{As Figure \ref{fig:disk-avg-abu-spec}, except at $90^{\circ}$S$-66^{\circ}$S (left) and $36^{\circ}$S$-12^{\circ}$S (right), assuming a global wet adiabat. Thick green and yellow lines plot the best-fitting H$_2$S and NH$_3$ profiles, respectively. The thick red line in the brightness spectra correspond to the best-fitting profiles. The gray box indicates the range of possible cloud top pressures and H$_2$S abundances detected by \citet{Irwin2019a} over the south polar cap, showing that H$_2$S is only detectable in our model at 100\% or greater relative humidity.}
\label{fig:sim-abu-wet}
\end{figure}

\begin{figure}
\centering
  \includegraphics[width=0.75\linewidth]{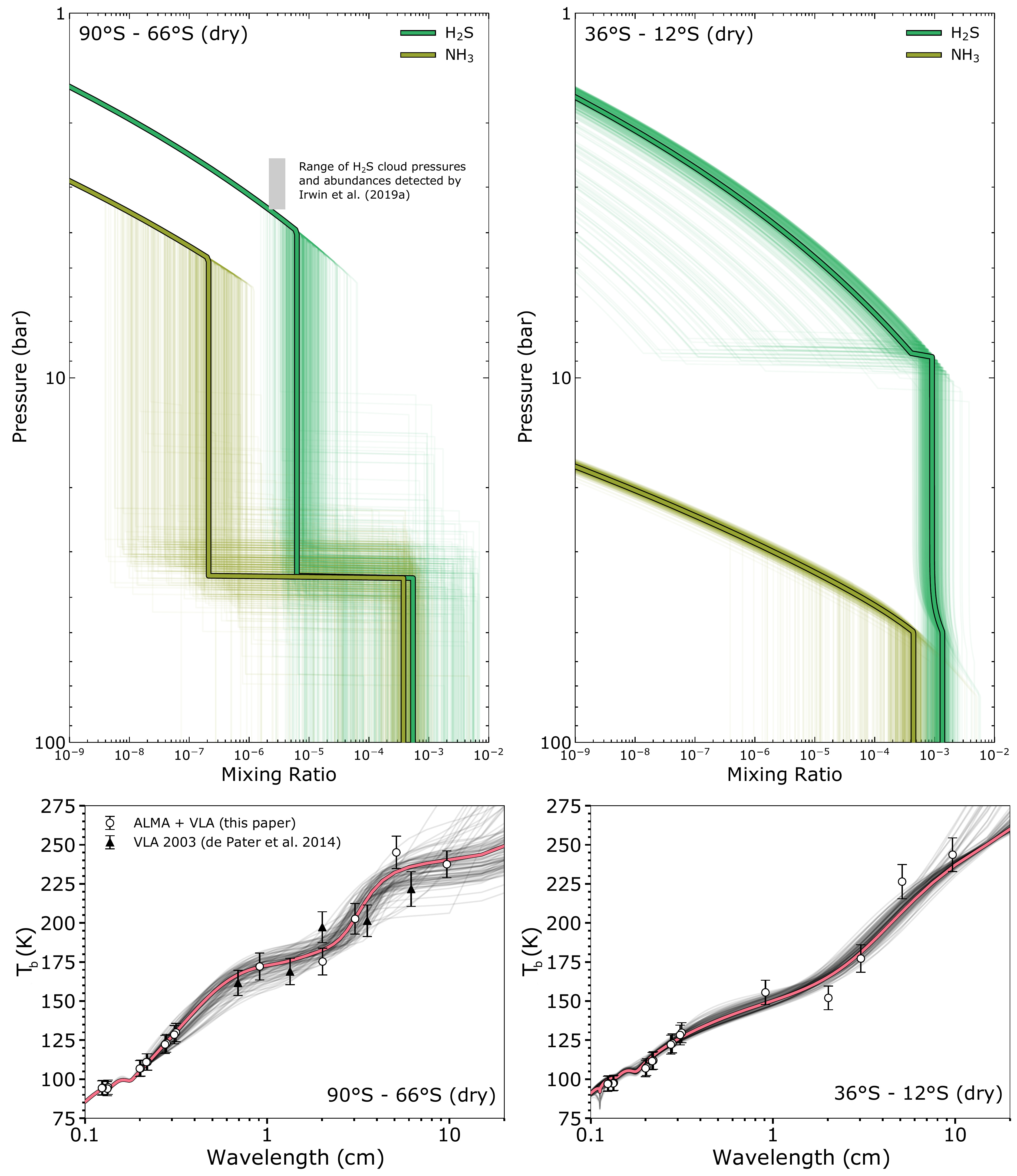}
\caption{As Fig \ref{fig:sim-abu-wet}, except assuming a global dry adiabat.}
\label{fig:sim-abu-dry}
\end{figure}

Figures \ref{fig:sim-abu-wet} and \ref{fig:sim-abu-dry} plot the best fitting abundance profiles for H$_2$S and NH$_3$ between $36^{\circ}$S$-12^{\circ}$S and $90^{\circ}$S$-66^{\circ}$S assuming wet and dry adiabat temperature profiles, respectively. Five hundred random profiles sampled from the burned-in posterior are also plotted, representing the range of allowable profiles. Brightness temperatures are also plotted, comparing one hundered random retrieved model spectra and data, showing an agreeable fit. A comparison to south polar 2003 VLA data are overplotted as well, showing good agreement. The assumption of a wet adiabat thermal profile does not produce H$_2$S abundances large enough to be detected by \citet{Irwin2019a} (see Fig. \ref{fig:sim-abu-wet}) unless H$_2$S is supersaturated. For the dry thermal profile, the H$_2$S-ice cloud forms at higher altitudes and larger H$_2$S abundances are possible. However, H$_2$S needs to be fully saturated or supersaturated to reach detectable upper atmosphere abundances.

\subsection{Globally-Varying Profiles - Mitigating Systematic Uncertainty}

One shortcoming of the analysis in the previous section is that the calibration error masks the evident brightness variations across the disk. In this section, we simultaneously model the atmospheric properties matching both the observed brightness temperature within the $36^{\circ}$S$-12^{\circ}$S latitudinal band, and the observed temperature differences between $36^{\circ}$S$-12^{\circ}$S and other latitudes. This approach reveals the cause of spatial trends in the brightness temperature. The new likelihood function to maximize is:

\begin{equation}
\text{ln } p(T_b|\nu, \theta, \sigma) = -0.5 \sum_{v_n} \left[\left(\frac{T_{b,n} - T_{b,m}}{\sigma_n}\right)^2 + \log(2 \pi \sigma_n^2)\right] - 0.5 \sum_{v_n} \left[\left(\frac{\Delta T_{b,n} - \Delta T_{b,m}}{\sigma_{\Delta, n}}\right)^2 + -\log(2 \pi \sigma^2_{\Delta, n})\right]
\end{equation}

The first sum on the right-hand side is the same as equation (1) and represents the fit to the observed brightness temperature at $36^{\circ}$S$-12^{\circ}$S with errors dominated by systematic calibration. The second sum represents the brightness temperature difference between $36^{\circ}$S$-12^{\circ}$S and the other latitudinal band:

\begin{equation}
\Delta T_b = T_{\text{$36^{\circ}$S$-12^{\circ}$S}} - T_{\text{other latitude}}.
\end{equation}

Errors in the temperature variation between latitudes at the same wavelength are mainly due to random fluctuations. The uncertainty $\sigma_{\Delta}$ is:

\begin{equation}
\sigma_{\Delta} = (\Delta T_b \cdot 0.05)^2 + \sigma_{RMS_1}^2 + \sigma_{RMS_2}^2 + \sigma_{PSF}^2
\end{equation}

The first term on the right-hand side is the difference in brightness temperature between the reference and modeled latitude band times the 5$\%$ calibration error. $\sigma_{RMS}$ is the random error on the sky divided by the square root of the number of beams that fit within the reference and modeled latitude bands. 

The deep abundances of H$_2$S, NH$_3$, and CH$_4$ are allowed to vary for each pair of latitudes. Moreover, H$_2$S may either subsaturate or supersaturate. H$_2$S may be supersaturated up to a pressure $P_{\text{ss}}$. At altitudes shallower than $P_{\text{ss}}$, the H$_2$S profile follows the saturation curve with 100\% relative humidity. At the south polar cap, the abundances of H$_2$S and NH$_3$ at pressures shallower than $P_{\text{mix}}$ are also allowed to vary, as in the previous section. We assume no PH$_3$ and fully intermediate H$_2$ at all latitudes as we do not expect to be sensitive to these parameters given our prior findings (Table \ref{table:lat-sim-dry-results}).

\begin{sidewaysfigure}
    \centering
    \includegraphics[width=\textwidth]{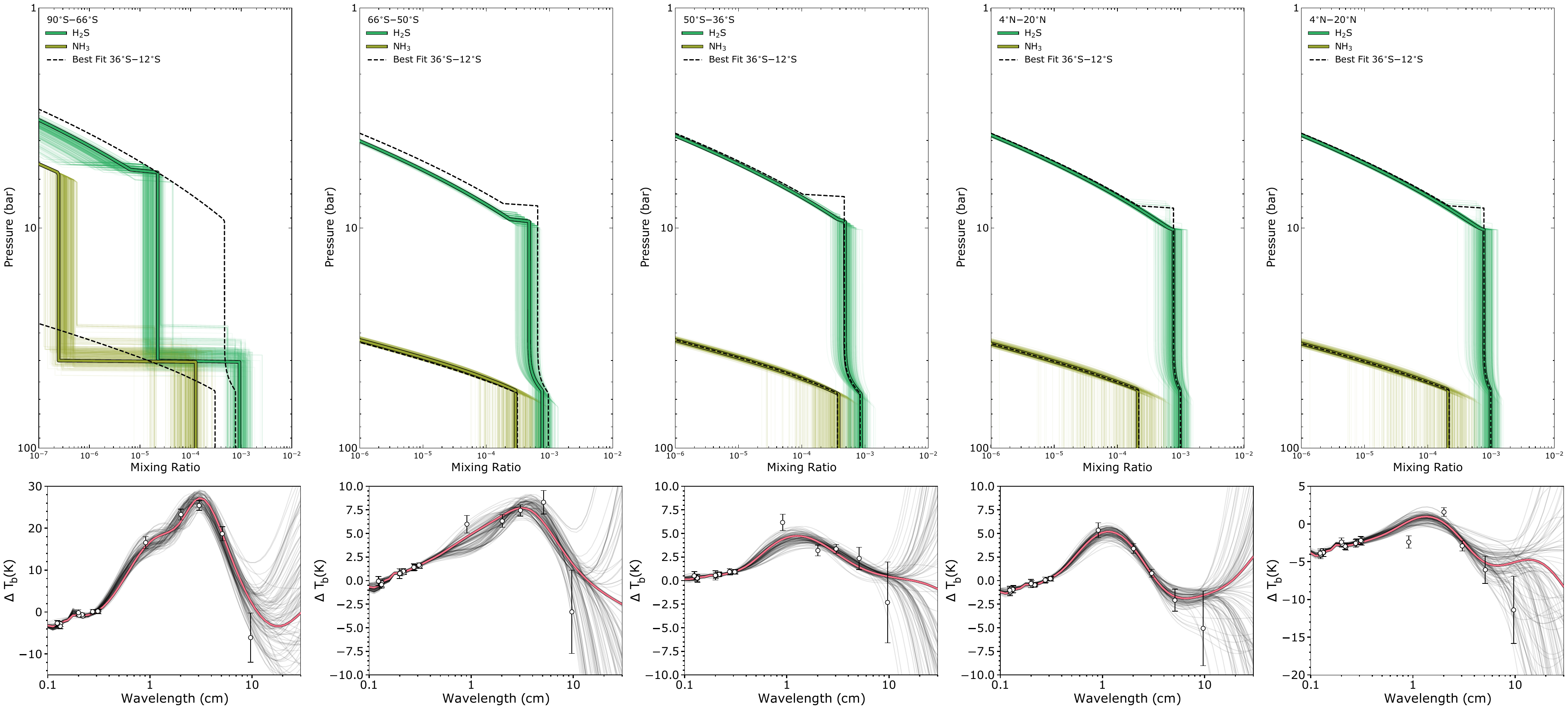}
    \caption{Retrieved abundance profiles (top) and residual brightness temperature spectra (bottom) across Neptune's disk going northward from left to right, assuming a global wet adiabat. Colors are as Figure \ref{fig:sim-abu-dry} and the dashed black lines denote the best-fitting $36^{\circ}$S$-12^{\circ}$S model from the simultaneous fit. Note that these profiles for $36^{\circ}$S$-12^{\circ}$S vary from panel-to-panel as its model parameters are retrieved independently for each simultaneous fit with the given latitude band. However, all are consistent to within the retrieved uncertainty. Good fits to the observed and modeled temperature variations are obtained. The brightness temperature variations, defined as $\Delta T_b$ in Eq. (5)., are plotted in the bottom panels, with the best-fit model plotted in red and 100 random models in black. Positive values indicate that the modeled latitude band is warmer than $36^{\circ}$S$-12^{\circ}$S at that wavelength.}
    \label{fig:lat-all-rms-abu-spec-res-wet}
\end{sidewaysfigure}

\begin{sidewaysfigure}
    \centering
    \includegraphics[width=\textwidth]{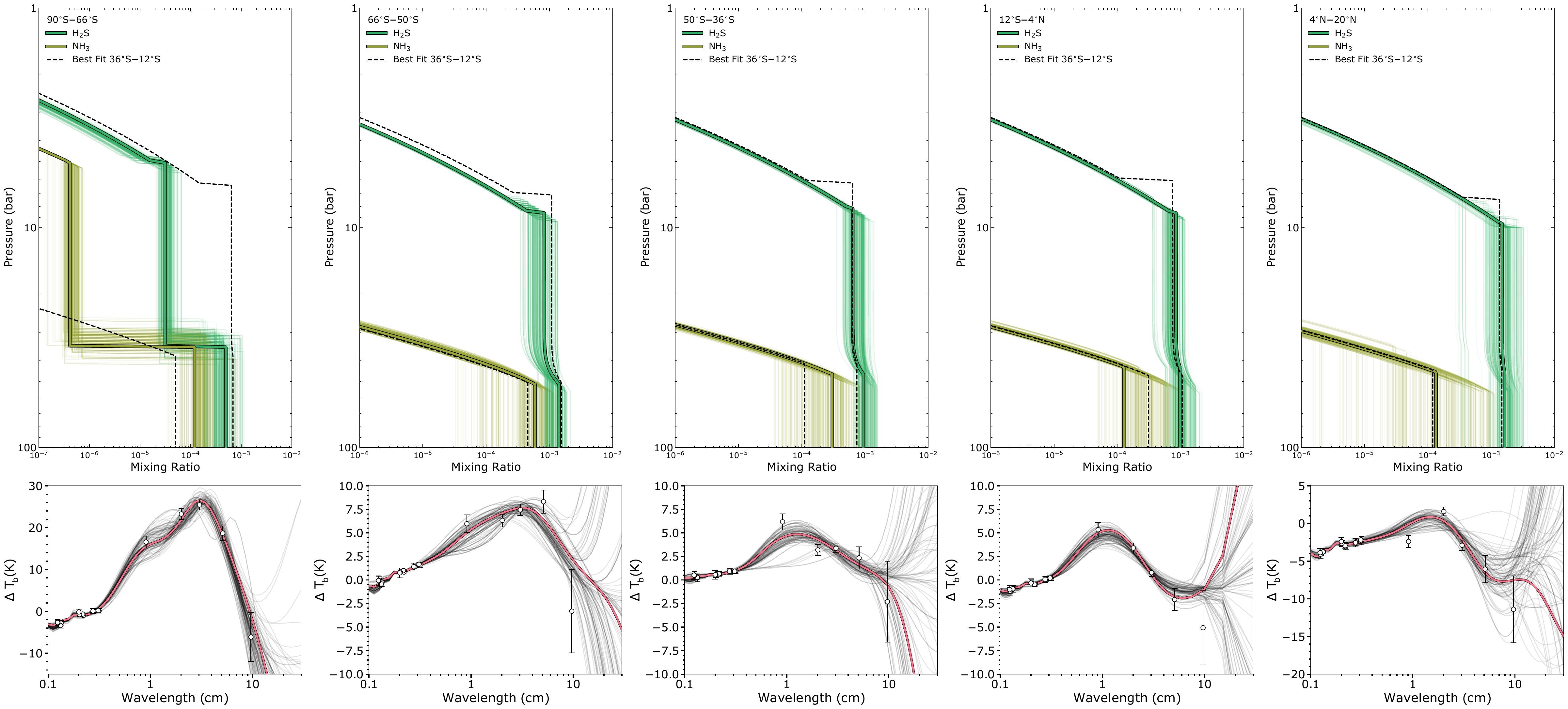}
    \caption{As Fig. \ref{fig:lat-all-rms-abu-spec-res-wet}, but assuming a global dry adiabat.}
    \label{fig:lat-all-rms-abu-spec-res-dry}
\end{sidewaysfigure}

Tables \ref{table:lat-all-results-wet-rms} and \ref{table:lat-all-results-dry-rms} list the best-fitting values for 16/50/84th uncertainties for constrained parameters assuming a global wet and dry adiabat, respectively. Figures \ref{fig:lat-all-rms-abu-spec-res-wet} and \ref{fig:lat-all-rms-abu-spec-res-dry} plot randomly-sampled and the best-fitting retrieved trace gas profiles and residual spectra $\Delta T$. We obtain good agreement between the observed and modeled temperature differences. Moreover, the retrieved parameters for the $36^{\circ}$S$-12^{\circ}$S band are consistent between model runs to within uncertainties, indicating this approach is self-consistent.  

\begin{sidewaysfigure}
\centering
  \includegraphics[width=\textwidth]{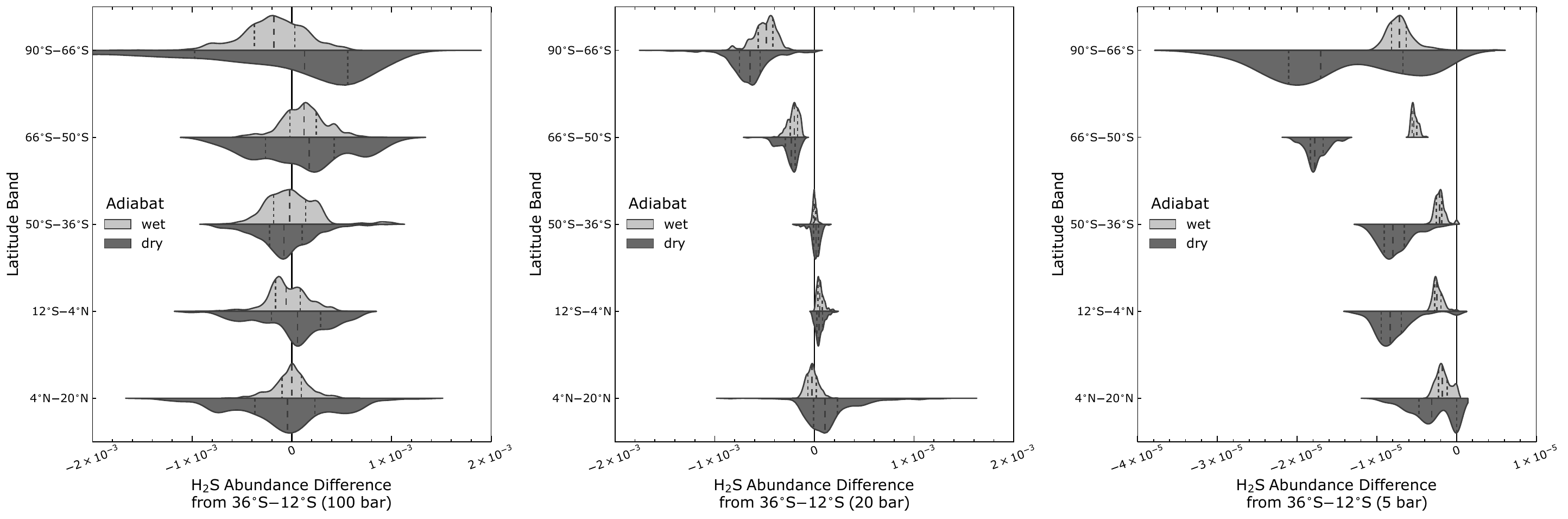}
\caption{Violin plots showing the probability density of the H$_2$S mole fraction for different latitude bands relative to values retrieved for the $36^{\circ}$S$-12^{\circ}$S band in the simultaneous fit. Negative values mean the amount of H$_2$S is higher at $36^{\circ}$S$-12^{\circ}$S. The three panels correspond to three altitudes: 100 bar (left), 20 bar (middle), and 5 bar (right). Distributions are split between the wet and dry thermal models, shaded light and dark respectively. Dashed lines show the median and upper and lower quartile values.}
\label{fig:h2s_violin}
\end{sidewaysfigure}

These results show global variability in Neptune's NH$_3$ and H$_2$S abundances. NH$_3$, although overall very much depleted due to the formation of NH$_4$SH, is enriched at the south polar cap above $P_{\text{mix}}$ relative to the rest of the planet. A globally uniform H$_2$S abundance below NH$_4$SH formation is consistent with our results. However, above NH$_4$SH formation, the H$_2$S abundance decreases poleward, mirroring brightness temperature increases poleward (Fig. \ref{fig:vla-maps}). Differences in the H$_2$S relative humidity cause the apparent brightness variations in the short-wavelength maps, which probe the H$_2$S-ice cloud and are coldest / darkest at $36^{\circ}$S$-12^{\circ}$S and $4^{\circ}$N$-20^{\circ}$N. These regions are also where supersaturated H$_2$S is needed in order to fit the observed temperature variations. This discussion is visually summarized in Figure \ref{fig:h2s_violin}, which plot the probability density of the retrieved H$_2$S mixing ratio in each latitude band relative to that obtained for $36^{\circ}$S$-12^{\circ}$S in the simultaneous fit. These plots demonstrate the magnitude and significance of these variations at the three pressure regimes discussed: below NH$_4$SH formation (100 bar), above H$_2$S-ice formation (5 bar), and in-between (20 bar). In the Appendix, we show that latitudinal variations in the thermal profile can only explain the observations if the temperature variations are localized to the H$_2$S-ice formation pressures and are on the order of $\sim 8$K or greater.

\begin{table}
\centering
\begin{tabular}{|c|c|c|c|c|c|}
\hline
Latitude Bin: & $90^{\circ}$S$-66^{\circ}$S & $66^{\circ}$S$-50^{\circ}$S & $50^{\circ}$S$-36^{\circ}$S & $12^{\circ}$S$-4^{\circ}$N & $4^{\circ}$N$-20^{\circ}$N\\
\hline
Free Parameter & \multicolumn{5}{ c| }{Retrieved values - global wet adiabat} \\
\hline
H$_2$S below NH$_4$SH & $9.5^{+2.5}_{-2.3}\times10^{-4}$ & $7.7^{+2.1}_{-1.3}\times10^{-4}$ & $8.0^{+1.3}_{-1.6}\times10^{-4}$ & $8.5^{+1.4}_{-1.3}\times10^{-4}$ & $9.2^{+2.4}_{-1.7}\times10^{-4}$ \\
NH$_3$ below NH$_4$SH & $0.9^{+0.5}_{-0.4}\times10^{-4}$ & $2.7^{+2.2}_{-1.1}\times10^{-4}$ & $3.1^{+1.3}_{-1.5}\times10^{-4}$ & $2.6^{+1.0}_{-1.2}\times10^{-4}$ & $1.7^{+1.6}_{-1.3}\times10^{-4}$ \\
CH$_4$ & $\leq0.5\times10^{-2}$ & $\leq0.6\times10^{-2}$  & $\leq0.6\times10^{-2}$  & $\leq0.8\times10^{-2}$ & $\leq0.5\times10^{-2}$ \\
H$_2$S $H_{\text{rel}}$ (\%)& $18^{+13}_{-11}$ & $49^{+5}_{-3}$ & $80^{+5}_{-5}$ & $76^{+8}_{-3}$ & $\geq70$ \\
$P_{\text{ss}}$ (bar) & not supersat. & not supersat. & not supersat. & not supersat. & $\geq 7.9$  \\
$P_{\text{mix}}$ (bar) & $38^{+3}_{-3}$ & --- & --- & --- & --- \\
H$_2$S above $P_{\text{mix}}$ & $1.4^{+0.6}_{-0.3}\times10^{-5}$ & --- & --- & --- & --- \\
NH$_3$ above $P_{\text{mix}}$ & $9.0^{+9.0}_{-3.9}\times10^{-7}$ & --- & --- & --- & --- \\

\hline
\hline
Free Parameter & \multicolumn{5}{ c| }{Corresponding $36^{\circ}$S$-12^{\circ}$S retrieved values} \\
\hline
H$_2$S below NH$_4$SH & $7.9^{+1.1}_{-0.9}\times10^{-4}$ & $9.3^{+1.5}_{-1.6}\times10^{-4}$ & $7.5^{+2.1}_{-1.3}\times10^{-4}$ & $8.1^{+1.7}_{-1.3}\times10^{-4}$ & $9.1^{+1.9}_{-1.3}\times10^{-4}$ \\
NH$_3$ below NH$_4$SH & $2.9^{+1.1}_{-1.4}\times10^{-4}$ & $2.1^{+1.9}_{-1.3}\times10^{-4}$ & $2.7^{+1.8}_{-1.3}\times10^{-4}$ & $2.8^{+1.6}_{-1.4}\times10^{-4}$ & $1.3^{+1.4}_{-1.0}\times10^{-4}$ \\
CH$_4$ & $\leq4.0\times10^{-2}$ & $\leq1.3\times10^{-2}$ & $\leq0.9\times10^{-2}$  & $\leq1.3\times10^{-2}$ & $\leq0.8\times10^{-2}$ \\
$P_{\text{ss}}$ (bar) &  $\geq 6.4$ & $\geq7.3$ & $7.2^{+0.3}_{-0.4}$ & $6.9^{+0.3}_{-0.3}$ & $8.2^{+0.5}_{-0.4}$  \\
H$_2$S $H_{\text{rel}}$ (\%) &  $\geq 48$ & $\geq 90$ & not subsat. & not subsat. & not subsat.  \\
$\sigma_{PSF}$ (\%) & $\leq 6$ & $\leq 1$ & $\leq 2$ & $\leq 2$ & $\leq 6$ \\
\hline
\end{tabular}
\caption{MCMC retrieved parameters for the simultaneous latitudinal band fitting, matching the observed brightness temperature variations at a given latitude band compared to that at $36^{\circ}$S$-12^{\circ}$S, assuming a global wet adiabat. A column in the top table lists the values of retrieved parameters at the given latitude band, while the corresponding column in the bottom table lists those obtained at $36^{\circ}$S$-12^{\circ}$S in the simultaneous fit (eq. 4). Free parameters marked with a $`-'$ were not allowed to vary at that latitude band. `not supersat.' and `not subsat.' means that supersaturated and subsaturated H$_2$S were not obtained in any final retrievals, respectively. The listed values are the 50th percentile of the posterior distributions with the error bars corresponding to the 16th and 84th percentiles, i.e.  the 1$\sigma$ uncertainties. Parameters for which the 16th and 84th percentiles vary by more than one order of magnitude have the 2.5th/97.5th percentile value listed instead, representing the 2$\sigma$ lower/upper limit. If both supersatured and subsaturated H$_2$S are possible in a latitude band, the 2$\sigma$ limit is listed for both $P_{\text{ss}}$ and H$_2$S $H_{\text{rel}}$.}
\label{table:lat-all-results-wet-rms}
\end{table}

\begin{table}
\centering
\begin{tabular}{|c|c|c|c|c|c|}
\hline
Latitude Bin: & $90^{\circ}$S$-66^{\circ}$S & $66^{\circ}$S$-50^{\circ}$S & $50^{\circ}$S$-36^{\circ}$S & $12^{\circ}$S$-4^{\circ}$N & $4^{\circ}$N$-20^{\circ}$N\\
\hline
Free Parameter & \multicolumn{5}{ c| }{Retrieved values - global dry adiabat} \\
\hline
H$_2$S below NH$_4$SH & $\leq5.7\times10^{-3}$ & $1.2^{+0.4}_{-0.3}\times10^{-3}$ & $1.0^{+0.3}_{-0.1}\times10^{-3}$ & $1.0^{+0.5}_{-0.2}\times10^{-3}$ & $1.9^{+0.5}_{-0.5}\times10^{-3}$ \\
NH$_3$ below NH$_4$SH & $1.7^{+0.9}_{-0.8}\times10^{-4}$ & $4.8^{+3.3}_{-3.5}\times10^{-4}$ & $2.2^{+2.3}_{-1.3}\times10^{-4}$ & $2.8^{+3.3}_{-1.5}\times10^{-4}$ & $\leq9.5\times10^{-4}$ \\
CH$_4$ & $\leq3.1\times10^{-2}$ & $\leq5.7\times10^{-2}$ & $\leq6.0\times10^{-2}$& $\leq1.5\times10^{-2}$ & $\leq2.3\times10^{-2}$ \\
H$_2$S $H_{\text{rel}}$ (\%)& $44^{+9}_{-11}$ & $53^{+4}_{-3}$ & $78^{+6}_{-4}$ & $78^{+5}_{-4}$ & $\geq61$ \\
$P_{\text{ss}}$ (bar) & not supersat. & not supersat. & not supersat. & not supersat. & $\geq 8.3$  \\
$P_{\text{mix}}$ (bar) & $34^{+3}_{-3}$ & --- & --- & --- & --- \\
H$_2$S above $P_{\text{mix}}$ & $\leq4.5\times10^{-4}$ & --- & --- & --- & --- \\
NH$_3$ above $P_{\text{mix}}$ & $\leq4.1\times10^{-6}$ & --- & --- & --- & --- \\

\hline
\hline
Free Parameter & \multicolumn{5}{ c| }{Corresponding $36^{\circ}$S$-12^{\circ}$S retrieved values} \\
\hline
H$_2$S below NH$_4$SH & $7.5^{+2.5}_{-1.7}\times10^{-4}$ & $1.4^{+0.4}_{-0.5}\times10^{-3}$ & $1.1^{+0.4}_{-0.4}\times10^{-3}$ & $9.8^{+2.2}_{-2.3}\times10^{-4}$ & $1.8^{+0.5}_{-0.4}\times10^{-3}$ \\
NH$_3$ below NH$_4$SH & $0.4^{+1.4}_{-0.2}\times10^{-4}$ & $3.2^{+3.2}_{-1.8}\times10^{-4}$ & $1.2^{+3.4}_{-0.6}\times10^{-4}$ & $3.0^{+4.8}_{-0.8}\times10^{-4}$ & $3.3^{+4.0}_{-2.3}\times10^{-4}$\\
CH$_4$ & $\leq5.3\times10^{-2}$ & $\leq1.6\times10^{-2}$ & $\leq6.7\times10^{-2}$  & $\leq1.5\times10^{-2}$ & $\leq2.7\times10^{-2}$ \\
$P_{\text{ss}}$ (bar) &  $6.2^{+1.0}_{-0.4}$ & $6.9^{+0.7}_{-0.3}$ & $\geq5.9$ & $6.1^{+0.3}_{-0.2}$ & $\geq6.8$   \\
H$_2$S $H_{\text{rel}}$ (\%) & not. subsat & not. subsat. & $\geq88$ & not subsat. & $\geq79$  \\
$\sigma_{PSF}$ (\%) & $\leq 4$ & $\leq 2$ & $\leq 1$ & $\leq 2$ & $\leq 9$ \\
\hline
\end{tabular}
\caption{As Table \ref{table:lat-all-results-wet-rms}, but assuming a global dry adiabat.}
\label{table:lat-all-results-dry-rms}
\end{table}

\begin{table}
\centering
\begin{tabular}{|c|c|c|c|c|c|c|}
\hline
Latitude Bin: & $90^{\circ}$S$-66^{\circ}$S & $66^{\circ}$S$-50^{\circ}$S & $50^{\circ}$S$-36^{\circ}$S & $12^{\circ}$S$-4^{\circ}$N & $4^{\circ}$N$-20^{\circ}$N & Total\\
\hline
Wet Adiabat DIC & 177 & 152 & 145 & 134 & 173 & 781  \\
Dry Adiabat DIC & 151 & 134 & 130 & 116 & 155 & 689 \\
Difference & +26 & +18 & +15 & +18 & +18 & +95\\
\hline
\end{tabular}
\caption{DIC values for each latitude band comparison per global temperature profile. Differences larger than 10 statistically favor the model with the lower DIC score, in this case the global dry adbaiat.}
\label{table:dic}
\end{table}

A measure of the improvement in the fit between two models is the Deviance Information Criterion (DIC, \citet{Spiegelhalter2002}). DIC is defined as:

\begin{equation}
\text{DIC} = \overline{D(\mathbf{\theta})} + \frac{1}{2}\text{Var}(D(\mathbf{\theta}))
\end{equation}

$\mathbf{\theta}$ are the free parameters, $\overline{D(\mathbf{\theta})}$ is the mean deviance of the retrieved parameters, which in this case is the average of the log-likelihood \textit{emcee} probabilities, defined in Eq. (4). The second term on the right hand side is one-half the variance of these probabilities.

Information criteria like DIC are typically used when analyzing MCMC results in order to select models which provide a good fit to the data (the first term) and minimize the variance in the retrieved probabilities (the second term) resulting from model complexity. The difference between two DIC values can be used to determine the better model assuming the parameters roughly follow a Gaussian distribution. Differences greater than 10 favor the model with the lower DIC score.

Table \ref{table:dic} lists the DIC score for the global wet and dry adiabat models at each latitude band. The dry adiabat is statistically favored everywhere, apart from $50^{\circ}$S$-36^{\circ}$S where neither thermal profile is favored. {We additionally show in Section 2 of the Appendix that when the lapse rate is allowed to vary with the trace gas abundances, the resulting retrieved temperature-pressure profiles follow a dry adiabat more closely than a wet adiabat.

\section{Discussion}

Retrievals of Neptune's microwave spectrum from radiative transfer modeling require an atmosphere dominant in H$_2$S over NH$_3$. In every retrieval, all NH$_3$ is taken up by the formation of the NH$_4$SH cloud around 50 bar, leaving H$_2$S to persist and condense at higher altitudes. The global thermal profile has a moderate impact on the produced H$_2$S and NH$_3$ profiles. Below all cloud formation, we obtain H$_2$S abundances of $8.7^{+2.6}_{-1.7}\times10^{-4}$ ($37.3^{+15.7}_{-10.3}\times$ protosolar) for a wet adiabat and $1.3^{+0.4}_{-0.3}\times10^{-3}$ ($53.8^{+18.9}_{-13.4}\times$ protosolar) for a dry adiabat. We are less sensitive to the deep NH$_3$ abundance as all contribution functions probe at and above the NH$_4$SH cloud. However, we do find a clear positive correlation between the H$_2$S and NH$_3$ abundances: more deep NH$_3$ requires more deep H$_2$S. We obtain loose bounds on the NH$_3$ abundance: $ 2.5^{+2.7}_{-2.3}\times10^{-4}$ ($2.1^{+2.3}_{-1.9}\times$ protosolar) for a wet adiabat and $ 4.7^{+2.5}_{-3.7}\times10^{-4}$ ($3.9^{+2.1}_{-3.1}\times$ protosolar) for a dry adiabat. Models fitting to the brightness temperature variations between latitude bands statistically favor a globally uniform thermal profile following a dry adiabat.

Using the H$_2$S/NH$_3$ ratio as a proxy for the S/N ratio, we completely rule out observable S/N ratios less than unity. This is in agreement with earlier radio work on Neptune, although we obtain better constraints on these abundances \citep{dePater1991, DeBoer1996, dePater2014, LuszczCook2013, Tollefson2019}. Sub-unity S/N is also ruled out on Uranus, suggesting some commonality in the formation history of the ice giants \citep{Gulkis1978, dePater1991, dePater2018b, Molter2020}. However, their true bulk S/N ratio may be much smaller than observed or assumed, due to the potential loss of ammonia in an ionic/superionic water ocean at 20 GPa \citep{Atreya2019}. Indeed, if CO is uplifted into Neptune troposphere, it implies that the water content and O/H ratio is several hundred times protosolar \citep{LuszczCook2013b}. An additional way to explain the absence of N is the preferential delivery of volatiles onto planetesimals via hydrated clathrates \citep{Hersant2004}. In the cold environment of the outer protoplanetary disk, the N$_2$ to NH$_3$ fraction is roughly 10, while S is primarily in the form of H$_2$S. Both H$_2$S and NH$_3$ are readily trapped in hydrated clathrates while N$_2$ gas is not, meaning the majority of available N will not be swept into the ice giants. The clathrated hydrates hypothesis also implies that the O/H ratio within Uranus and Neptune is at least 100 times protosolar.

The differences in brightness temperature between latitude regions reveal prevalent global variations in the H$_2$S and NH$_3$ profiles. Brightness temperature variations between latitude bands were modeled by comparing the retrieved spectra to profiles simultaneously fit to the $36^{\circ}$S$-12^{\circ}$S region. This approach limits the calibration error, placing tight constraints on how the trace gas profiles vary latitudinally.

At the south pole, the data can be fit with a downwelling atmosphere, where air depleted in H$_2$S and NH$_3$ species subsides down to a mixing pressure, $P_{\text{mix}}$, below which the atmosphere becomes well mixed and the trace species equal their deep values. The best-fit retrieved H$_2$S and NH$_3$ mixing ratios at altitudes above $P_{\text{mix}}$ are $4.4\times10^{-5}$ and $4.1\times10^{-7}$, respectively, for a dry adiabat. This is in broad agreement with south polar models of 2003 VLA data presented in de Pater et al. (2014), whose best fitting depleted model required $3.5\times10^{-5}$ parts H$_2$S and $1.2\times10^{-8}$ parts NH$_3$ above the NH$_4$SH cloud.

Below NH$_4$SH formation, our results are consistent with homogenous NH$_3$ and H$_2$S. Clear latitudinal trends in the H$_2$S profile, however, must be present at higher altitudes. Just above NH$_4$SH cloud formation, the H$_2$S abundance is highest between around the equator and mid-latitudes and diminishes poleward.

\citet{Irwin2019a} tentatively detected $1-3$ ppm H$_2$S at Neptune's cloud tops between $2.5-3.5$ bar. Their detection is more robust near the south pole than the equator and both their retrieved cloud top pressures and abundances increase toward the south pole. Our results assuming a wet adiabat globally are not consistent with their findings (for instance, see the gray rectangles in Fig. \ref{fig:sim-abu-wet}). The wet thermal profile is about 7K cooler at 3 bar than the \citet{Irwin2019a} profile. Only our dry adiabat models at the south pole allow H$_2$S gas abundances close to the values found by \citet{Irwin2019a}, as warmer temperatures push the H$_2$S cloud base to higher altitudes.

Supersaturated H$_2$S between $36^{\circ}$S$-12^{\circ}$S and possibly between $4^{\circ}$N$-20^{\circ}$N is required to obtain good fits to the observed brightness temperature variations if the thermal profile is fixed globally. Numerous near-infrared works favor a two-layer cloud/haze structure featuring a shallow tenuous upper haze and an opaque cloud deck between $2-4$ bar \citep{Irwin2014, LuszczCook2016, Molter2019}. This lower cloud is presumably H$_2$S-ice, supporting our finding. The ratio of the partial pressure of the condensate in a supersaturated model to the partial pressure of condensate following its saturation curve, $\phi$, is a few hundred percent in our H$_2$S supersaturation model. Similar degrees of supersaturation are expected within ammonia plumes on Jupiter \citep{dePater2019}. However, on Earth $\phi$ does not exceed $10\%$ for water \citep{Young1993}. This is because the timescale of cloud formation is very quick in the presence of cloud condensation nuclei (CCN). Little is known about the amount of CCNs available on the gas giants. But, theoretical calculations by \citet{Moses1992} of homogeneous, heterogeneous, and ion-induced nucleation rates produce large values and ranges for $\phi$, depending on the species. They find $\phi \geq 3-1000$ in order for methane and photolyzed hydrocarbon aerosols to form in Neptune's stratosphere. Neptune may lack substantive amounts of CCN conducive to aerosol formation. Future work should consider both the effect of cloud micro-physics and dynamics on the trace gas distributions within the ice giants. 

The above discussion is synthesized in Figure \ref{fig:h2s-global}, which shows one potential global H$_2$S profile. For each latitudinal band, we assume the best-fitting H$_2$S abundances and relative humidities from Table \ref{table:lat-all-results-dry-rms}. Broadly, these meridional trends in gas abundances are related to the global circulation patterns of the planet.

Warm temperatures within Neptune's south polar cap are detected at mid-infrared and radio wavelengths. In the mid-infrared, \citep{Hammel2007} observe a warm south pole explained by prevalent ethane and methane emission. \citep{Orton2007} further argue that seasonal warming at the south polar cap may explain the excess of these molecules in the stratosphere; the warm temperatures would overcome the cold-trapping of methane below its condensation point at $\sim1$ bar and allow methane to escape upward, diffuse globally, and form ethane via photolysis. \citep{Orton2007} also argue that rising air would be expected given the unexpected high abundances of these species. However, warm temperatures relative to the rest of the planet seen in the mid-infrared thermal emission \citep{Fletcher2014} and a lack of widespread, persistent cloud coverage in the near-infrared (e.g., \citet{dePater2014}) are more indicative of dry subsiding air. Radio observations, including these, observe high brightness temperatures at Neptune's south polar cap \citep{LuszczCook2013, dePater2014, Tollefson2019}, which we also interpret as dry, low-opacity, subsiding air. Cloud coverage surrounding the south polar cap and faint, distinct near-infrared clouds located near, but not at, the south pole might be indicative of vigorous convection \citep{LuszczCook2010}, analogous to Saturn's polar activity \citep{Dyudina2008, Fletcher2008}. This vigorous convection may be a mechanism to explain all observations across the wavelength spectrum.

The mid-latitudes, defined loosely as between $50^{\circ}$S$-15^{\circ}$S and northward of $15^{\circ}$N, are where Neptune's brightest and strongest methane cloud activity is observed in the near-infrared \citep{Martin2012, Fitzpatrick2014, dePater2014, Tollefson2018}. In contrast, the equatorial region is nearly featureless. In addition, the mid-latitudes are colder than the equator and south pole in the mid-infrared \citep{Conrath1998, Fletcher2014}. Combining these observations, a global circulation pattern is inferred at altitudes shallower than $\sim1$ bar (where the near- and mid-infrared probe): cold, enriched methane air rises at the mid-latitudes and travels to the equator and poles, where the methane-depleted air subsides and warms via adiabatic compression. However, this picture is complicated by the relative excess of gaseous CH$_4$ at the equator, which is more consistent with rising air \citep{Karkoschka2011, Tollefson2018, Irwin2019b}. We show in this paper that H$_2$S is most abundant at the equator and southern mid-latitudes, in line with this meridional trend in CH$_4$. At altitudes below methane condensation ($\sim1$ bar), the aforementioned circulation scheme is, thus, at odds with the retrieved CH$_4$ and H$_2$S abundances. Therefore, a more complicated picture of global circulation on Neptune is needed to explain each multi-wavelength observation. \citet{Fletcher2020} synthesize decades worth of analysis on the ice giants and argue that vertically-stacked circulation cells are necessary to bridge the observed patterns above and below 1 bar on both Uranus and Neptune.

\begin{figure}
    \centering
    \includegraphics[width=0.75\textwidth]{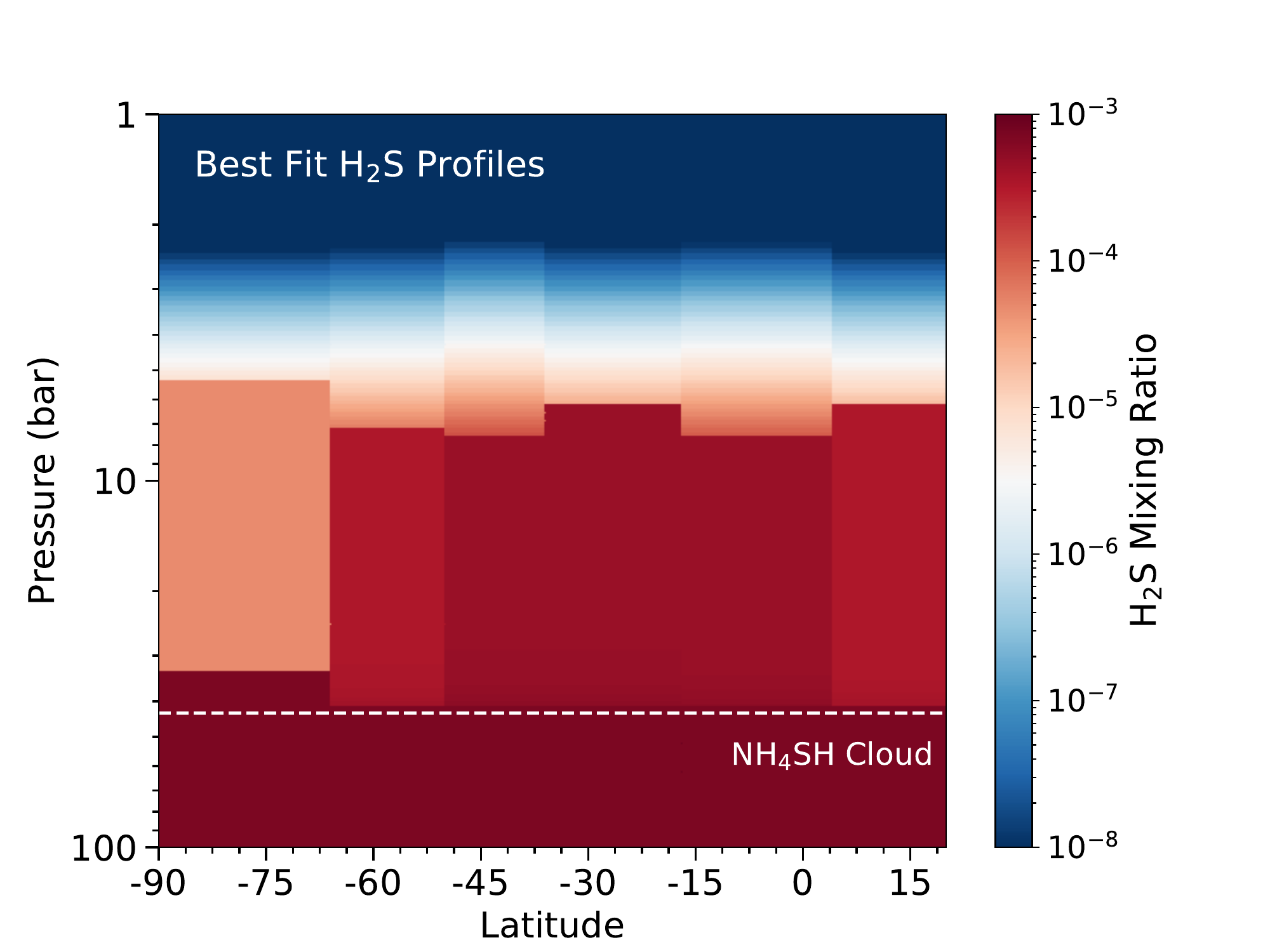}
    \caption{The best-fitting global H$_2$S profile from MCMC retrievals.}
    \label{fig:h2s-global}
\end{figure}

\section{Conclusions}

We observed Neptune at radio wavelengths with the Very Large Array in five bands between $0.9-9.7$ cm from $1-2$ September, 2015. The longitude-smeared maps reveal brightness variations across Neptune's disk. These variations are alternating dark and bright latitudinal bands. Dark (bright) bands are consistent with high (low) opacity sources and cold (warm) brightness temperatures. We model Neptune's brightness temperature distribution using the radiative transfer code Radio-BEAR coupled to MCMC, varying the abundance profiles of Neptune's condensibles to obtain best-fits and retrievals to the observed microwave spectra. Models are fit to data from both the VLA presented in this work and $1-3$ mm ALMA maps from \citet{Tollefson2019}. Combined, these data probe from 1 bar down to $>$50 bar, where the NH$_4$SH cloud forms.

\begin{itemize}
\item The assumed thermal profile has a moderate impact on the retrieved H$_2$S, NH$_3$, and CH$_4$ profiles. A global dry adiabat is preferred over a wet adiabat as a warmer thermal profile is statistically favored in models fitting brightness temperature variations across the disk. All below results are given for the dry adiabat models.

\item The abundances of H$_2$S and NH$_3$ below all cloud formation are  $1.3^{+0.4}_{-0.3}\times10^{-3}$ ($53.8^{+18.9}_{-13.4}\times$ protosolar) and $4.7^{+2.5}_{-3.7}\times10^{-4}$ ($3.9^{+2.1}_{-3.1}\times$ protosolar). 

\item The abundance of CH$_4$ is unconstrained due to its limited impact on the microwave opacity longward of 1 cm. In the mm, the CH$_4$ contribution to the modeled spectra competes with other opacity sources, namely PH$_3$, the H$_2$ state, and the H$_2$S relative humidity, dulling its signal in the retrievals.

\item The downwelling south polar cap ($90^{\circ}$S$-66^{\circ}$S) is consistent with depleted H$_2$S and NH$_3$ down to a mixing pressure $P_{\text{mix}}$ around the NH$_4$SH cloud formation: $\sim33$ bar. Only when H$_2$S is fully saturated or supersaturated can retrieved H$_2$S abundances exceed 1 ppm at 3.5 bar at the south polar cap, as measured by \citet{Irwin2019a}. Only our dry adiabat model can produce values at pressures close to those detected. 

\item The observed brightness temperature variations between latitude bins are consistent with decreased H$_2$S above the NH$_4$SH cloud deck moving away from $36^{\circ}$S$-12^{\circ}$S.

\item The observed brightness temperature variations are more consistent with models supersaturating H$_2$S between $36^{\circ}$S$-12^{\circ}$S, and possibly $4^{\circ}$N$-20^{\circ}$N, than with models following thermochemical equilibrium with some subsaturation. Models which do not allow H$_2$S to supersaturate require latitudinal variations in the kinetic temperature on the order of 8K or greater to fit the data. These variations in the thermal profile are localized to H$_2$S-ice formation pressures. Future work is needed to decouple the effects of temperature and gas opacity on Neptune's radio spectrum.

\item We find no spectral evidence of PH$_3$. Our retrievals are consistent with a PH$_3$ $2\sigma$ upper limit of 0.4 ppm ($\leq 0.1\times$ protosolar).

\item We cannot constrain the ortho/para H$_2$ fraction anywhere on Neptune.
\end{itemize}

Advances in radio astronomy are critical for further constraining the abundance of condensibles in the ice giants. A next generation VLA (ngVLA) would improve the resolution ten-fold at 20 cm if the longest 1000 km baselines are utilized \citep{dePater2018, Selina2018}. The resulting high-quality data at long wavelengths would permit a strong constraint on the NH$_3$ abundance beneath the NH$_4$SH layer and perhaps even on H$_2$O. In addition, a five-fold improvement in the sensitivity would mean vastly shorter integration times to achieve an equivalent RMS to this work, meaning zonal variations can be detected over a similar observing period and global maps can be obtained. \textit{Juno}'s big advantage over ground-based radio observations of Jupiter is that it flies beneath the synchrotron radiation belts that dominate longward of 6 cm. This is not an issue on the ice giants. Moreover, the resolution and sensitivity between the ngVLA and a MWR-like instrument on a potential future Neptune orbiter are quite similar \citep{dePater2020}.

The decision to include an MWR-equivalent on a direct mission to the ice giants will have to weigh whether it competes too heavily with the ground-based capabilities of an optimally running ngVLA. Whatever the choice, it is clear that the ngVLA or a direct mission to the planet are required to improve our maps of Neptune's trace species and elucidate theories regarding the environment in which Neptune formed and evolved.


\acknowledgments

This paper makes use of the following VLA data from program VLA/15A-118. The National Radio Astronomy Observatory (NRAO) is a facility of NSF operated under cooperative agreement by Associated Universities, Inc. VLA data used in this report are available from the NRAO Science Data Archive
at https://archive.nrao.edu/archive/advquery.jsp.

This paper makes use of the following ALMA data: ADS/JAO.ALMA$\#$2016.1.00859.S. ALMA is a partnership of ESO (representing its member states), NSF (USA) and NINS (Japan), together with NRC (Canada), MOST and ASIAA (Taiwan), and KASI (Republic of Korea), in cooperation with the Republic of Chile. The Joint ALMA Observatory is operated by ESO, AUI/NRAO and NAOJ. The National Radio Astronomy Observatory is a facility of the National Science Foundation operated under cooperative agreement by Associated Universities, Inc.

This research was supported by the National Science Foundation, NSF Grant AST-1615004 to the University of California, Berkeley.

E. Molter was supported in part by NRAO Student Observing Support grant \#SOSPA6-006.

\appendix

\setcounter{table}{0}
\renewcommand{\thetable}{A\arabic{table}}
\setcounter{figure}{0}
\renewcommand{\thefigure}{A\arabic{figure}}

\subsection{Globally Uniform Temperature Profiles}

In this section, we address the assumption that the temperature profile is globally uniform at a given pressure level. We first show that this assumption leads to reasonable wind speeds according to the thermal wind equation, then demonstrate that not allowing the trace gas abundances to vary meridionally can result in well-modeled spectra only for nonphysical temperature profiles.

We test the validity of the assumption that the meridional temperature gradient is zero for $P > 1$ bar using an order-of-magnitude comparison. The thermal wind equation relates the vertical wind shear to both the meridional gradient in temperature and composition:

\begin{equation}
2\Omega \sin \theta \frac{du}{dr} = -\frac{g}{r_{\text{eq}}} \left(\frac{1}{T}\frac{dT}{d\theta} + \frac{C}{1+Cq}\frac{dq}{d\theta} \right) 
\end{equation}

$\Omega$ is Neptune's rotation rate ($1\times10^{-4}$ rad s$^{-1}$), $\theta$ is the latitude (rad), $u$ is the zonal wind velocity (m s$^{-1}$), $r$ is altitude (m), $r_{\text{eq}}$ is Neptune's equatorial radius ($2.5\times10^{7}$ m), $g$ is Neptune's gravitational constant (10 m s$^{-2}$), $T$ is the temperature, $q$ is the molar mass fraction of CH$_4$ (the most prominent trace gas on Neptune), and $C = (1-\epsilon)/\epsilon$ where $\epsilon$ is the molar mass ratio of CH$_4$ compared to the ambient atmosphere. For the purpose of this exercise, we assume that the ambient atmosphere consists of H$_2$, He, and CH$_4$; H$_2$S and NH$_3$ have comparatively smaller abundances than CH$_4$ and are therefore less important to the compositional term. $\epsilon  \approx 7$ and $C \approx -1$. At the southern mid-latitudes, $\sin \theta \approx -0.5$ and the above equation becomes:

\begin{equation}
-1\times10^{-4} \frac{du}{dr} \approx 4\times10^{-7} \frac{1}{1 - q}\frac{dq}{d\theta} 
\end{equation}

From \citet{Karkoschka2011}, $q$ is about 0.1 at Neptune’s south pole ($\leq 1 - 2\%$ mixing ratio) and $0.2$ toward the southern mid-latitudes and equator ($\geq 2 - 4\%$). So $dq/d(\theta) \sim +0.1$ at the mid-latitudes for $1 \leq P \leq 3$ bar. Thus, the right hand side is about $5 \times 10^{-8}$ and $du/dr \approx -0.5$ m s$^{-1}$ km$^{-1}$. The sign and magnitude of this estimate is similar to that of \citet{Tollefson2018} who observed vertical wind shear at Neptune's equator by tracking near-infrared clouds. Therefore, the thermal wind equation alone cannot rule out the assumption of constant meridional temperatures. 

Since the vertical wind shear is relatively unconstrained, especially below 1 bar, meridional kinetic temperature variations are not precluded from a thermal wind analysis. To test this, we estimate what temperature profile is needed to model the observed residual brightness temperatures. We fix the trace gas abundances globally to equal the nominal model ($30\times$S H$_2$S, CH$_4$, H$_2$O and $1\times$S NH$_3$, with no PH$_3$, ‘intermediate’ H$_2$, and 100$\%$ relative humidity at all cloud formation). Fig \ref{fig:appendix-res-temp-lat} plots the observed and modeled temperature variations versus wavelength, meaning a dry adiabat model for $36^{\circ}$S-$12^{\circ}$S is subtracted from the model temperature at each other latitude bin. The temperature profiles used to create these models are plotted in \ref{fig:appendix-res-temp-prof}. To fit the residual temperatures, two adjustments to the dry adiabatic lapse rate are required. If the kinetic temperature were the sole cause of the brightness temperature variations, then substantial local variations on the order of 8K or greater are required. Interestingly, these peak around pressures where the H$_2$S-ice cloud forms ($\sim10$ bar), suggesting that the natural explanation for meridional brightness variability is due to changes in the H$_2$S profile altering the microwave opacity.

\begin{figure}
    \centering
    \includegraphics[width=0.95\textwidth]{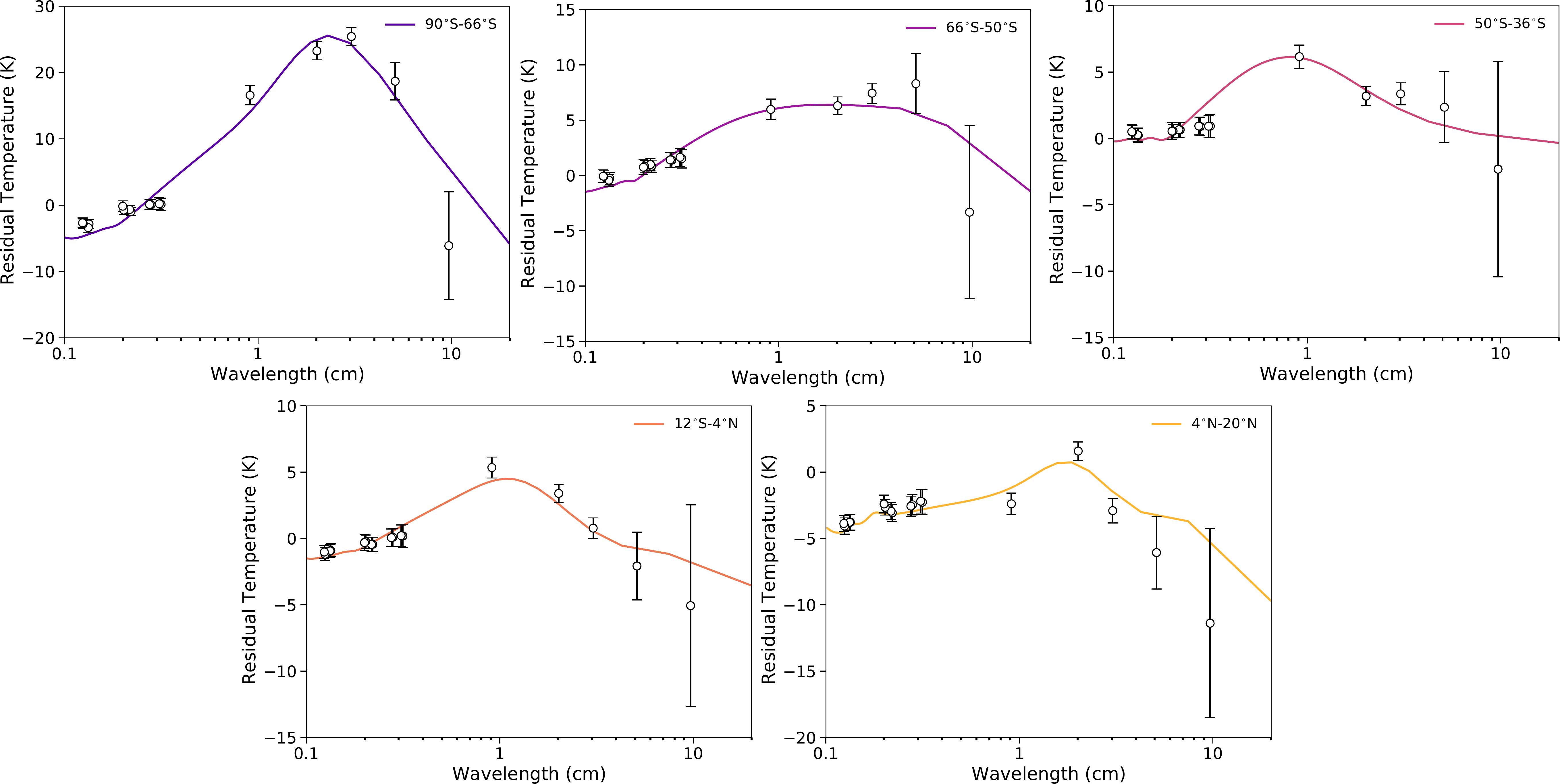}
    \caption{Residual temperatures as a function of latitude, where the observed temperature at $36^{\circ}$S-$12^{\circ}$S is subtracted from the temperature at the given latitude band. Colored lines are example model fits to the data assuming a nominal abundance profile and the same colored temperature profile given in Fig. \ref{fig:appendix-res-temp-prof}.}
    \label{fig:appendix-res-temp-lat}
\end{figure}

\begin{figure}
    \centering
    \includegraphics[width=0.95\textwidth]{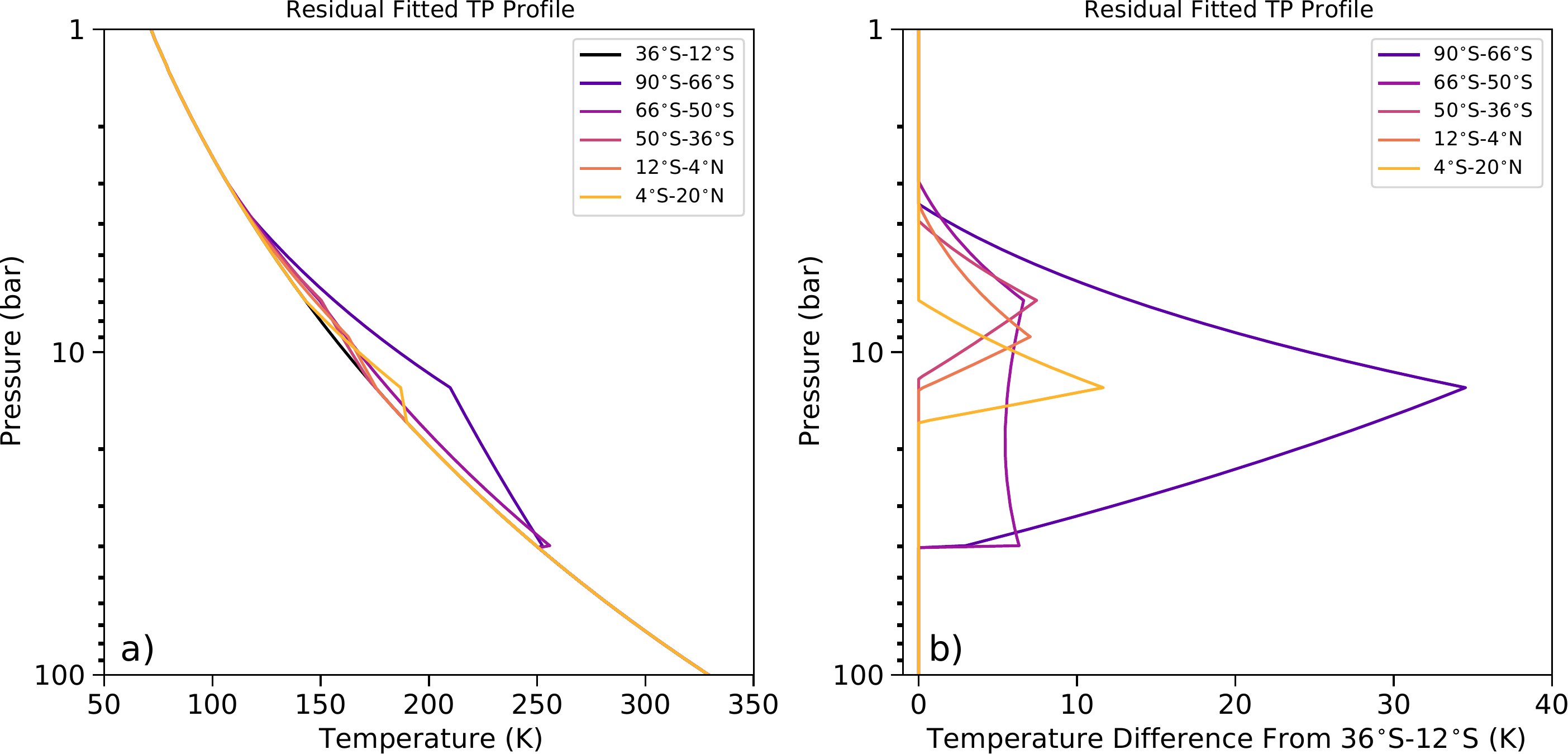}
    \caption{a) Temperature profiles as a function of latitude (colored lines) producing the residual spectra fits plotted in Fig. \ref{fig:appendix-res-temp-lat}. b) Residual temperature profile as a function of latitude (colored lines), where the  $36^{\circ}$S-$12^{\circ}$S dry adiabat profile is subtracted from the temperature at the given latitude band.}
    \label{fig:appendix-res-temp-prof}
\end{figure}

\subsection{Varying Temperature and Trace Gas Abundances with MCMC}

In the following, we consider a combination of kinetic temperature and trace gas abundance variations that may explain the observed brightness temperature latitudinal differences. To vary the kinetic temperature in the forward RT model, we allow the lapse rate to vary:

\begin{equation}
    \Gamma = -\frac{dT}{dz}
\end{equation}

Assuming hydrostatic equilibrium and the ideal gas law:

\begin{equation}
    \Gamma = \rho g \frac{dT}{dP} = \frac{PMg}{RT}  \frac{dT}{dP},
\end{equation}

where $R$ is the ideal gas constant, $g$ is the gravitational constant for Neptune, and for each layer $P$ is its pressure, $T$ its kinetic temperature, and $M$ its molecular mass. 

The molecular mass of the layer, $M$, is set by the mole fraction abundance of H$_2$, He, and CH$_4$ where [He]/[H$_2$] = 18$\%$, in line with infrared results from \citet{Burgdorf2003}, and [CH$_4$] = 1.44$\%$ (30$\times$ protosolar) at altitudes deeper than its condensation pressure. Equation (4) is integrated and solved for $T(P)$. For this work, $T(P)$ is altered by varying the lapse rate at altitudes shallower than a prescribed boundary pressure. At altitudes deeper than this boundary pressure, $T(P)$ is set to the dry adiabat profile used throughout this paper. We run three models fitting $\Delta T_b$ between $66^{\circ}$S-$50^{\circ}$S and $36^{\circ}$S-$12^{\circ}$S. Model 1 sets the boundary pressure at the onset of H$_2$S-ice formation and does not allow H$_2$S to supersaturate. Model 2 sets the boundary pressure at the onset of NH$_4$SH formation and does not allow H$_2$S to supersaturate. Model 3 sets the boundary pressure at the onset of NH$_4$SH formation and \textit{does} allow H$_2$S to supersaturate.  Both the lapse rate and condensible profiles are varied independently between latitude bands. 

Figures \ref{fig:66S-tempvar-profs} and \ref{fig:66S-tempvar-spec} plot the retrieved gas and temperature profiles and their corresponding residual spectra, respectively, for each of the three models. Table \ref{table:varytemp} lists the retrieved model parameters, showing that similar NH$_3$ and H$_2$S profiles are obtained regardless of the chosen boundary pressure, where the lapse rate begins to vary. Figure \ref{fig:chisq-tempvar} plots the $\chi^2$ goodness-of-fit as a function of wavelength for the best fitting spectra in each of the three models considered. Only when supersaturated is allowed (model 3) can a good fit be obtained at 0.9 cm. Allowing latitudinal temperature variations without H$_2$S supersaturation (models 1 and 2) can not explain all the observed data. In fact, the retrieved latitudinal differences in the kinetic temperature are consistent with zero, as shown in Figure \ref{fig:hist-tempvar}. While the 0.9 cm datum is just one point of evidence in favor of supersaturated H$_2$S, it can not be explained away by an error in calibration, as discussed in Section 4.3. Figure \ref{fig:66S-tempvar-profs} also shows that the retrieved temperature-pressure profiles are consistent with a dry adiabat while few retrievals follow the lapse rate of a wet adiabat, consistent with the findings in Section 4.2. We therefore conclude that latitudinal variations in the opacity due to the condensible species are the cause of Neptune's observed brightness temperature distribution, and the current fitting favors supersaturation of H$_2$S at the coldest latitudes.

\begin{sidewaysfigure}
\centering
  \includegraphics[width=\textwidth]{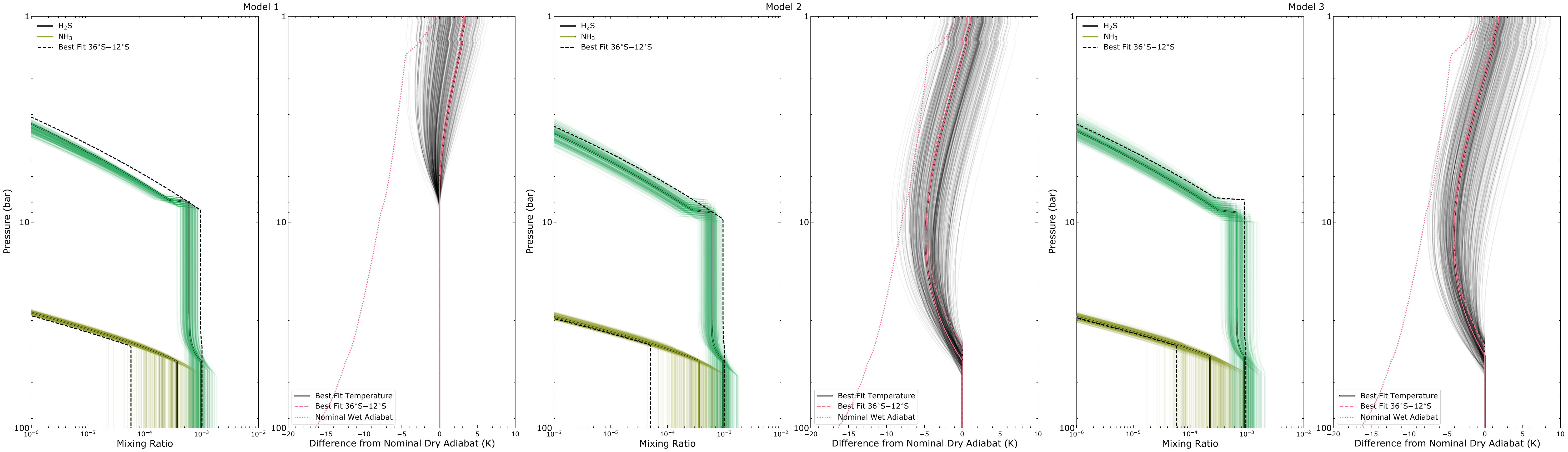}
\caption{Random retrieved abundance and temperature profiles compared to the nominal dry adiabat at $66^{\circ}$S-$50^{\circ}$S for the three types of models considered in this section. The best fitting profiles at $66^{\circ}$S-$50^{\circ}$S are given in bold colored lines while the best fitting profiles at $36^{\circ}$S-$12^{\circ}$S are given in dashed lines. The temperature profiles are differences from the nominal dry adiabat, with positive values meaning the retrieved model is warmer. The nominal wet adiabat is given in a dotted line for comparison and the difference in shape from the retrievals is due to forcing the temperature at 1 bar to match the \textit{Voyager} result.}
\label{fig:66S-tempvar-profs}
\end{sidewaysfigure}

\begin{figure}
\centering
  \includegraphics[width=\textwidth]{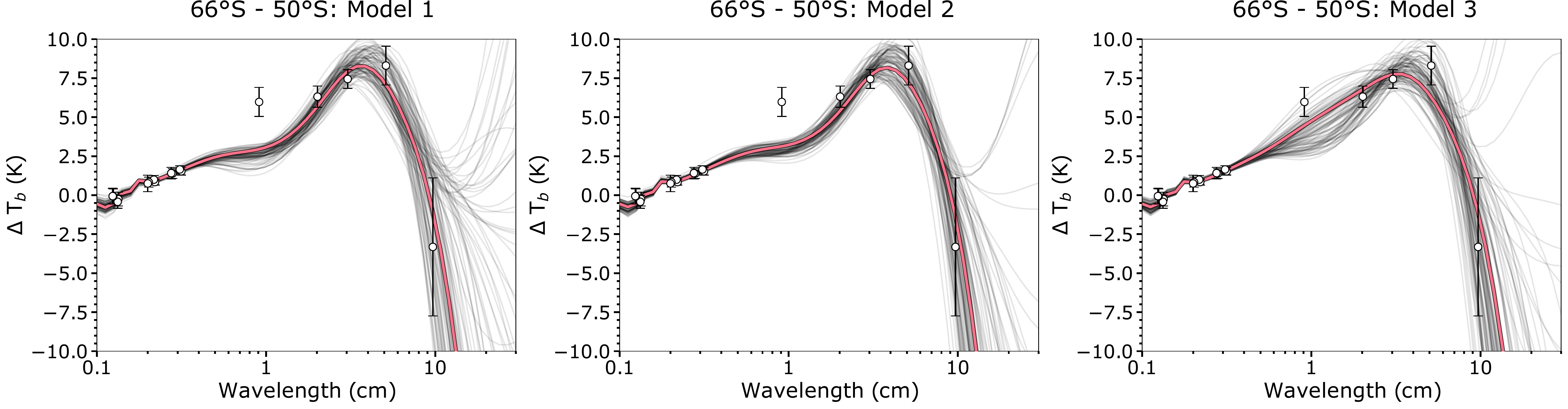}
\caption{Random retrieved model spectra fitting the observed $\Delta T_b$ between $66^{\circ}$S-$50^{\circ}$S and $36^{\circ}$S-$12^{\circ}$S for the three types of models considered in this section. Only model 3, which allows H$_2$S supersaturation at its ice cloud, provides a good fit to the data at 0.9 cm. }
\label{fig:66S-tempvar-spec}
\end{figure}

\begin{table}
\centering
\begin{tabular}{|c|c|c|}
\hline
Latitude Bin: & $66^{\circ}$S$-50^{\circ}$S & $36^{\circ}$S$-12^{\circ}$S \\
\hline
Free Parameter & \multicolumn{2}{ c| }{Retrieved values - Model 1} \\
\hline
H$_2$S below NH$_4$SH & $0.9^{+0.3}_{-0.2}\times10^{-3}$ & $1.1^{+0.3}_{-0.2}\times10^{-3}$ \\
NH$_3$ below NH$_4$SH & $2.6^{+2.8}_{-1.3}\times10^{-4}$ & $0.6^{+2.5}_{-0.3}\times10^{-4}$ \\
H$_2$S $H_{\text{rel}}$ ($\%$) & $48^{+3}_{-6}$& $91^{+7}_{-14}$ \\
$\Gamma$ (K/km) & $1.17^{+0.04}_{-0.06}$ & $1.17^{+0.05}_{-0.05}$ \\
\hline
Free Parameter & \multicolumn{2}{ c| }{Retrieved values - Model 2} \\
\hline
H$_2$S below NH$_4$SH & $1.0^{+0.2}_{-0.3}\times10^{-3}$ & $1.0^{+0.2}_{-0.2}\times10^{-3}$ \\
NH$_3$ below NH$_4$SH & $3.3^{+2.9}_{-1.8}\times10^{-4}$ & $0.4^{+1.0}_{-0.3}\times10^{-4}$ \\
H$_2$S $H_{\text{rel}}$ ($\%$) & $44^{+5}_{-6}$& $92^{+6}_{-13}$ \\
$\Gamma$ (K/km) & $1.11^{+0.03}_{-0.02}$ & $1.12^{+0.03}_{-0.03}$ \\
\hline
Free Parameter & \multicolumn{2}{ c| }{Retrieved values - Model 3} \\
\hline
H$_2$S below NH$_4$SH & $1.0^{+0.3}_{-0.3}\times10^{-3}$ & $1.1^{+0.3}_{-0.3}\times10^{-3}$ \\
NH$_3$ below NH$_4$SH & $2.5^{+2.7}_{-1.7}\times10^{-4}$ & $0.5^{+1.8}_{-0.4}\times10^{-4}$ \\
H$_2$S $H_{\text{rel}}$ ($\%$) & $51^{+4}_{-6}$ & $\geq72$ \\
$P_{ss}$ (bar) & --- & $\geq6.8$ \\
$\Gamma$ (K/km) & $1.11^{+0.02}_{-0.03}$ & $1.12^{+0.02}_{-0.04}$ \\
\hline
\end{tabular}
\caption{Retrieved parameters for each of the three models considered in this section.}
\label{table:varytemp}
\end{table}

\begin{figure}
\centering
  \includegraphics[width=0.5\textwidth]{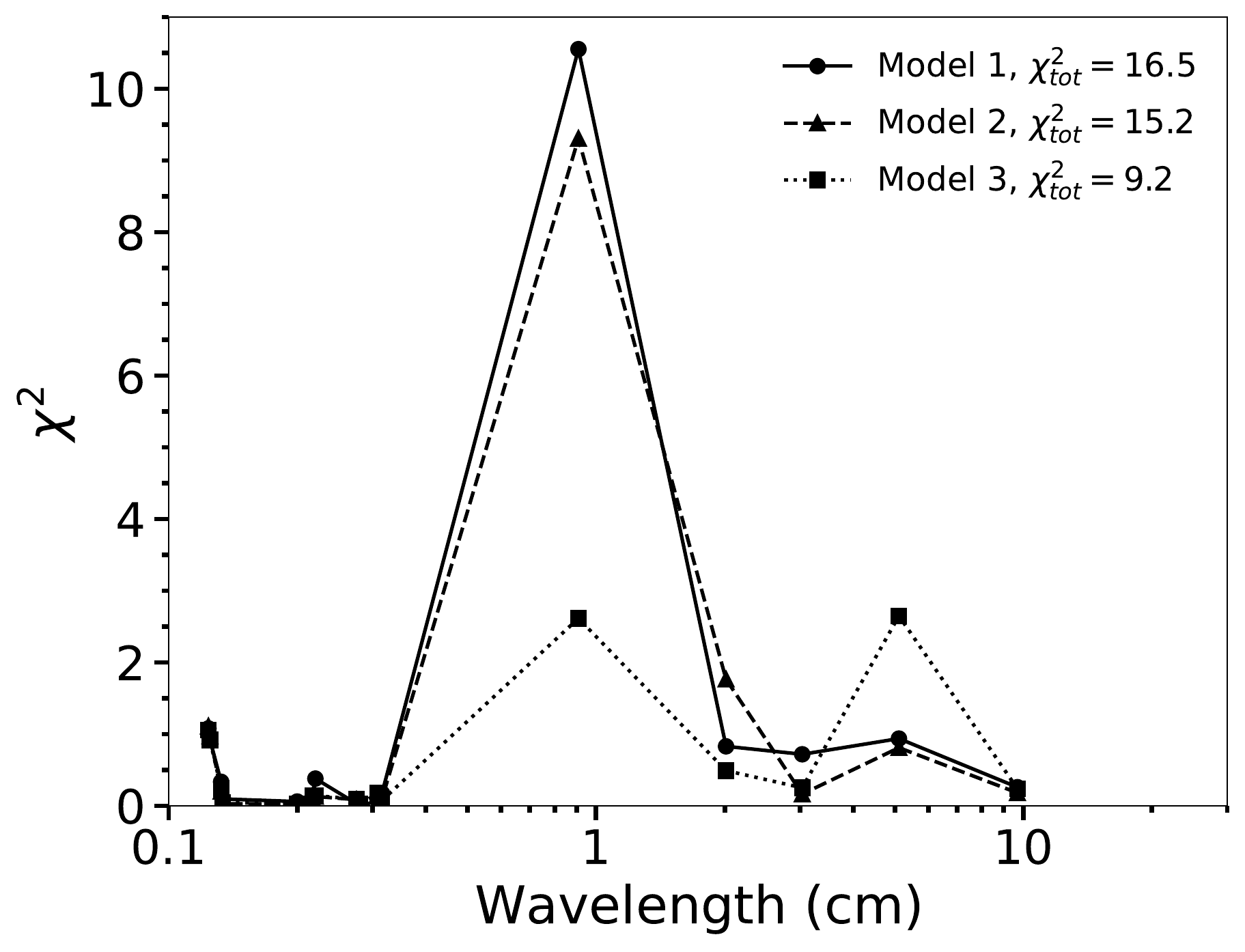}
\caption{$\chi^2$ goodness-of-fit versus wavelength corresponding to the best fitting residual spectra $\Delta T_b$ given in Fig. \ref{fig:66S-tempvar-spec}, showing that a good fit is achieved everywhere only when H$_2$S is allowed to supersaturate (Model 3). Varying temperature without allowing supersaturation (models 1 and 2) does not produce a good fit at the 0.9 cm data point. The total $\chi^2$ for each model is given in the plot legend.}
\label{fig:chisq-tempvar}
\end{figure}

\begin{figure}
\centering
  \includegraphics[width=0.5\textwidth]{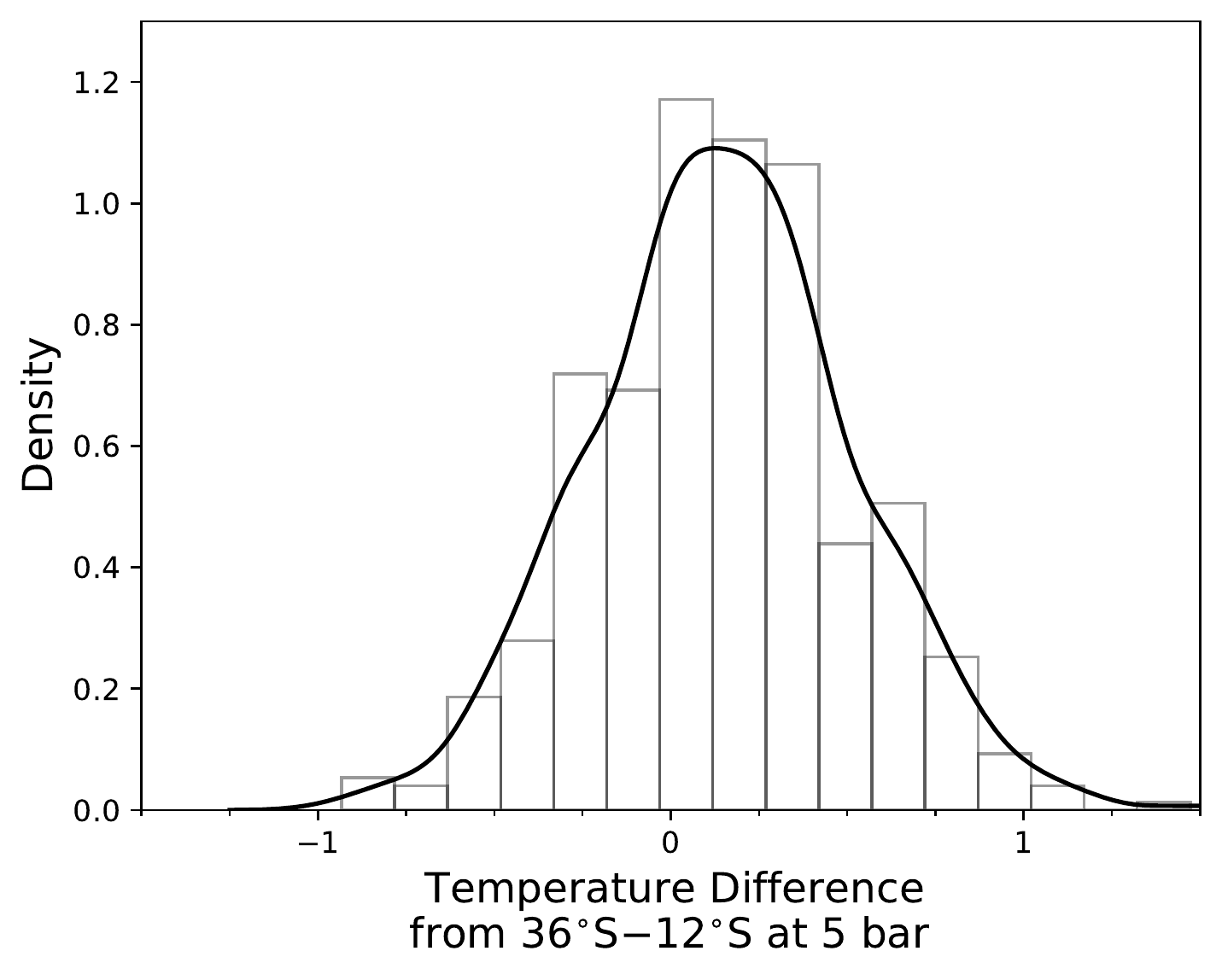}
\caption{Distribution of retrieved kinetic temperature differences between $66^{\circ}$S-$50^{\circ}$S and $36^{\circ}$S-$12^{\circ}$S at 5 bar for model 3. Differences are consistent with zero and no larger than 1 K. }
\label{fig:hist-tempvar}
\end{figure}

\pagebreak

\end{document}